\documentclass[pdflatex,sn-mathphys-num]{sn-jnl}

\usepackage{graphicx}%
\usepackage{multirow}%
\usepackage{amsmath,amssymb,amsfonts}%
\usepackage{amsthm}%
\usepackage{mathrsfs}%
\usepackage[title]{appendix}%
\usepackage{xcolor}%
\usepackage{textcomp}%
\usepackage{manyfoot}%
\usepackage{booktabs}%
\usepackage{algorithm}%
\usepackage{algorithmicx}%
\usepackage{algpseudocode}%
\usepackage{listings}%
\usepackage{hyperref}
\usepackage[capitalise]{cleveref}

\usepackage{chngcntr} 

\theoremstyle{thmstyleone}%
\theoremstyle{thmstyletwo}%

\theoremstyle{thmstylethree}%

\begin{document}

\title[Article Title]{Reinforcement Learning-based adaptive time-integration for nonsmooth dynamics}


\author*[1]{\fnm{David Michael} \sur{Riley}}\email{david.riley@ensta.fr}

\author[2]{\fnm{Alexandros} \sur{Stathas}}\email{alexandros.stathas@boku.ac.at}

\author[1]{\fnm{Diego} \sur{Guti\'errez-Oribio}}\email{diego.gutierrez@ensta.fr}

\author[1]{\fnm{Ioannis} \sur{Stefanou}}\email{ioannis.stefanou@ensta.fr}

\affil[1]{%
  \orgname{IMSIA, CNRS, EDF, ENSTA, Institut Polytechnique de Paris},%
  \orgaddress{\city{Palaiseau}, \postcode{91120}, \country{France}}%
}

\affil[2]{%
  \orgdiv{Institute of Structural Engineering Department of Landscape, Water and Infrastructure},%
  \orgname{~BOKU University},%
  \orgaddress{\street{~Peter-Jordan-Stra\ss e~82}, \postcode{1190}, \city{Wien}, \country{Austria}}%
}


\abstract{Numerical time integration is fundamental to simulating initial and boundary value problems across science, engineering, and even finance. Traditionally, time integration schemes require adaptive time-stepping to ensure computational speed and sufficient accuracy. Either directly or indirectly, these schemes require a certain degree of smoothness of the underlying dynamics. However, a significant number of problems in nature are nonsmooth. In this case, conventional schemes struggle to accurately integrate the solution or often fail to do so in a timely manner. In this work, we use an alternative approach based on Reinforcement Learning (RL) to select the optimal time step for any time integrator method, balancing computational speed and accuracy. Our RL approach learns to optimally adapt its time step through training, even in the presence of nonsmoothness. This capability makes it a prominent method for integrating a large spectrum of challenging dynamical systems.
To highlight its potential, we consider three model problems of increasing complexity. We begin with a nonlinear control method (sliding mode control), then proceed to an electrical circuit with a diode, and finally address a frictional instability problem that models a seismic fault with Coulomb friction. These examples collectively demonstrate the robustness of our strategy in handling nonsmooth, set-valued dynamics across diverse spatiotemporal scales—a notoriously challenging scenario for conventional numerical time integrators. Our results indicate that the RL-based adaptive integrator can learn an optimal time-integration strategy and achieve considerable speed-ups in relatively simple case scenarios, as well as a tenfold speed-up in the most challenging cases with spatiotemporal complex dynamics. These promising findings suggest that our approach can provide a novel and efficient alternative for time integration in various dynamical systems.}

\keywords{Reinforcement Learning, numerical time integration, adaptive time stepping, nonsmooth dynamics}



\maketitle

\section{Introduction}
\label{sec1}

Many problems in science, engineering, or finance exhibit sudden changes in their behavior, making it challenging for traditional methods to accurately predict and simulate their future states. These systems are described by evolution equations that lack regularity and are not analytic, rendering them inherently nonsmooth. The ubiquity of these types of systems is evident as they are observed across various disciplines such as mechanics~\cite{brogliato2016nonsmooth,acary2008numerical}, electrical circuits~\cite{acary2008numerical,bernardo2008piecewise}, biology~\cite{casey2006piecewise,bernardo2008piecewise} and control theory (e.g., sliding mode controllers/observers)~\cite{b:SMC_Fridman},  as well as in market models for finance~\cite{tramontana2010complicated} or energy~\cite{valencia2020non} to mention a few. 

Conventional time integration techniques, which resolve the time evolution of dynamical systems~\cite{hairer1993solving}, typically rely on constant time stepping. However, the accuracy and efficiency of such methods depend heavily on the chosen time step. While constant time stepping is straightforward, it can prove computationally intensive, especially for stiff systems or systems with multiple spatiotemporal scales, which are common in nonsmooth dynamics.

To address the computational inefficiency of constant time-stepping, adaptive integration methods have been developed~\cite{hairer1993solving}. There are generally two frameworks for undertaking adaptive time integration. The first is self-adaptive methods, such as adaptive Runge-Kutta schemes, which use internal information to adjust the time step ~\cite{hairer1993solving,press2007numerical}. An alternative approach is to estimate local errors by comparing solutions from one large time step with those from two smaller time steps, known as Richardson extrapolation~\cite{hairer1993solving}. While these methods have mathematical foundations, they rely on local approximations of the error, implying that they are neither exact nor global. To enhance these methods, research has introduced heuristic parameters to ensure that the time step neither changes too slowly nor too rapidly, which often requires careful empirical tuning~\cite{acary2009toward}.
Most importantly, smoothness is critical to the assumptions of conventional adaptive integrators. For stiff and nonsmooth systems, additional considerations must be taken into account, such as integrator order reduction~\cite {acary2008numerical,soderlind2002automatic} or loss of differentiability~\cite{pfau2007priori}. To address these phenomena in constrained mechanical systems, adaptive integrators have been proposed that aim to predict constraint switching~\cite {acary2008numerical,pfau2007priori}. However, in large systems, tracking constraint switching incurs additional computational overhead and a tremendous reduction in time step size. Consequently,  the benefits of this particular method are unclear~\cite{acary2008numerical}.

Yet another alternative is based on the mathematical theory of control~\cite{ogata2020modern}. These methodologies rely on designing a controller to choose the appropriate step size to achieve a given accuracy~\cite{gustafsson1991control,soderlind2002automatic,arevalo2020software}. Control-theoretic methods have generally focused on ensuring a smooth time-step evolution rather than computational speed. Moreover, they require distinct control gains tailored for either stiff and nonstiff ODEs~\cite{soderlind2002automatic}. Thus, control-based time-steppers are not well-suited to general dynamical systems.

Recently, advances in machine learning have led research to focus on replacing conventional time-integration methods for dynamical systems with data-driven approaches. A notable class of methods is neural ordinary differential equations (neural ODEs), in which ML is used to learn the mapping from the current state $x(t)$ to the state at time $ x(t+h)$~\cite{chen2018neural}. These methods do not rely on specific time-integration schemes, and their efficiency depends on how well the neural operator has learned the mapping. Recently, various advancements have been made in applying neural ODEs to stiff dynamical systems~\cite{kim2021stiff,fronk2025training,caldana2025neural} and even to differential-algebraic equations~\cite{koch2024learning}, but, critically, these advances rely on sufficient smoothness in the rates of the state and are therefore not applicable to nonsmooth dynamical systems. An alternative approach is to use existing time integration schemes, paired with ML, to learn the optimal time step for advancing the state forward in time, as done in~\cite{ouala2021learning}.In~\cite{ouala2021learning}, they apply ML to learn the coefficients of an explicit $q$-stage Runge–Kutta method. This optimal time step can be trained to achieve a desired balance between accuracy and computation time, while accounting for the system's intrinsic dynamics, which, in our case, are nonsmooth.

Combining deep learning with adaptive time stepping has shown significant improvement over conventional adaptive integrators. For instance,~\cite{liu2022hierarchical} applied 
a purely data-driven neural network for multiple time-scale dynamics. In this work, a small set of neural networks, each covering a different, fixed step size, is used to chain them together, allowing the simulation to leap forward by large increments before a classical numerical integrator takes over for the fine-scale details. Although effective on smooth or mildly chaotic systems, it is data-intensive, and its accuracy degrades when the dynamics include sharp events such as impacts or switching. More recently, an analogous technique was used to select the optimal time step from a predefined, discrete set (i.e., a discrete action space). The method was applied to chaotic, yet smoothly varying, ODEs using an explicit integrator~\cite{dellnitz2023efficient}, yielding very satisfactory results. Similarly, a variant of the same RL-based explicit adaptive time stepper, but with a continuous action space, was developed and applied to a fluid-solid particle simulation to improve computational speed while maintaining sufficient accuracy~\cite{han2021artificial, zhu2023ai}. More involved RL-based architectures have also been recently proposed. Based on an implicit integrator, ~\cite{dong2022adaptive,xu2023adaptive} employed two independent neural networks to determine the next time step. One of the neural networks was responsible for advancing in time, while the other was responsible for retreating in time if the implicit solver did not converge. They applied this forward-backward stepping strategy to simulate electrical circuits, yielding quantitatively accurate results, given that the underlying integrator was implicit and required nonlinear solver iterations. However, while this dual-network structure performs well on smooth problems, the forward-backward stepping strategy could be avoided, given the current potential of newer RL paradigms. Recently, these RL-based adaptive integrators have been applied to the 3-body problem in astrophysics, achieving sufficient accuracy, yet they were not competitive in computational time compared to commonly used adaptive integrators for such problems~\cite{ulibarrena2025reinforcement}.
More broadly, although all of the aforementioned approaches represent important achievements in the time integration of smooth numerical problems, they do not exploit the significant potential of Reinforcement Learning for nonsmooth dynamical systems, which is the focus of the current work.

It is worth highlighting that RL provides an effective framework for handling nonsmooth systems. This arises from RL's inherent ability to learn and adapt the time step to abrupt, severe state changes, characteristic of nonsmooth dynamical systems. Consequently, RL can generate optimal predictions for subsequent time-step sizes. Notice that conventional local rate measures, such as time derivatives of the state, are inadequate for nonsmooth systems, as these derivatives do not exist in such contexts (see the concept of subdifferential~\cite{rockafellar1997convex}). A prominent example illustrating this limitation is the stick-slip phenomenon encountered in frictional motion, which will be explored in~\cref{sec: example}.

In this paper, we develop an RL-based approach for adaptive time-stepping of nonsmooth dynamical systems using a single Truncated Quantile Critics (TQC) network, leveraging the robustness of TQC for continuous action spaces~\cite{kuznetsov2020controlling}. This single-network approach not only simplifies the architecture compared to the dual-network strategy but also provides a general implementation suitable for explicit and implicit integration schemes of nonsmooth dynamical systems. The extension of RL-based adaptive time-stepping approaches to nonsmooth dynamical systems enables the methodology to efficiently integrate systems with discontinuities and set-valued right-hand sides for the first time. To demonstrate its versatility, we present three model problems of increasing complexity: a first-order sliding-mode controller, an electrical circuit, and a frictional instability model simulating a seismic fault with Coulomb friction. Our findings suggest that the RL-based approach, combined with variational inequalities for solving the nonsmooth constrained systems~\cite{acary2008numerical}, establishes a general framework for learning efficient and accurate policies for adaptive time-stepping that can be applied across a diverse range of complex dynamical systems.

This paper is organized as follows. In~\cref{prob descrip}, the general form of the type of dynamical systems relevant to this paper is introduced.~\cref{sec: heuristic} briefly introduces the heuristic-based adaptive time stepper and a proportional integral (PI) based time stepper. Following suit,~\cref{RL: description} introduces the Reinforcement Learning algorithm employed, and, finally, the performance of the RL-based method is showcased with three case studies in~\cref{sect: RL perfomance}.

\section{Problem description}
\label{prob descrip}
To explore the potential of Reinforcement Learning for adaptive time-stepping policies, we investigate the general class of systems characterized by:
\begin{equation} \label{eq: dynamics}
    \mathcal{A} \dot{w} = F_s(t,w) + F_r(t,w),
\end{equation}
where $\mathcal{A} \in \mathbb{R}^{N \times N}$ is an invertible matrix, $w \in \mathbb{R}^{N}$ is the state vector that fully describes the condition of the system at a given time, $N$ is the number of degrees of freedom, and $\dot{\Box}=\frac{\partial}{\partial t}$ is the time derivative. The term $F_s(t,w) \in \mathbb{R}^{N}$ is a smooth function encompassing continuous dynamics such as internal forces and/or external fields, while $F_r(t,w) \in \mathbb{R}^{N}$ accounts for the nonsmooth terms arising from constraints that impose limits or conditions on the state variables. Many dynamical systems exhibit such nonsmooth constraints, including mechanical systems with friction, electrical circuits with transistors, financial markets with regulations that abruptly restrict trading options, and ecological systems where resource scarcity instantly limits population growth. This explicit distinction between smooth dynamics $F_s(t,w)$ and nonsmooth constraints $F_r(t,w)$ enables the systematic study and modeling of a broad range of constrained systems, encompassing holonomic constraints (state-based restrictions) and non-holonomic constraints (rate-based restrictions),

\subsection{Numerical constraint enforcement and integration}
Several numerical approaches exist for enforcing constraints in dynamical systems. The classical methods typically enforce constraints via penalty formulations or Lagrangian multipliers~\cite{crisfield1997nonlinear,Bauchau2007,Laulusa2007}. However, penalty formulations often lead to numerical stiffness or accuracy issues, particularly when very large penalty parameters are required to enforce constraints effectively. While more precise than penalty approaches, Lagrangian multiplier methods can increase computational complexity and introduce numerical difficulties related to solving saddle-point problems~\cite{franceschini2022reverse}. 

A convex-analytic alternative, adopted here, avoids both issues by geometrically defining the admissible constraint forces through a single closed convex set $\mathcal{C}$. This set encodes all admissible states or rates of the system. The constraint forces arise to ensure the system remains within $\mathcal{C}$, but their behavior depends on the nature of the constraint. For example, in an electrical circuit with a switch, current flows only when the switch is closed, creating an immediate change in the circuit's behavior—a nonsmooth transition where the constraint force (the resistive effect) instantaneously changes from infinite to finite. Conversely, in mechanical systems with friction, the constraint force belongs to a convex set defined by Coulomb's law, where the magnitude of the frictional force is bounded by the normal force multiplied by the friction coefficient. This force opposes motion within these bounds when sliding occurs, or exactly counteracts applied forces when static.  This geometric interpretation translates into a mathematical framework—a variational inequality—that unifies the treatment of diverse constraints. Efficient projection-based solvers leverage this framework to enforce constraints exactly, avoiding the stiffness of penalty methods and the saddle-point complexities of Lagrangian formulations~\cite{acary2008numerical,acary2018solving}.

Accurate numerical integration is often a critical factor in simulations of dynamical systems, particularly when nonsmooth behaviour is present. We employ the Bathe method~\cite{bathe2012insight,zhang2020improved}, which introduces numerical damping of high-frequency content while preserving low-frequency content and maintaining second-order temporal accuracy. However, the methodology applied in this paper can also be applied to alternative time-stepping methods.

\section{Heuristic-based Adaptive and PI time-stepping}
\label{sec: heuristic}
To evaluate the effectiveness of our Reinforcement Learning-based adaptive time-stepping policy, we first introduce the heuristic-based adaptive time-stepping method commonly used in numerical integration packages~\cite{SciPyDocs,shampine1997matlab}. This method will serve as a baseline for comparison. Another comparison could be with packages such as \textsc{MATLAB}/\texttt{solve\_ivp}, but these must employ an event-driven approach, which differs from the nonsmooth solvers we use here. A further comparison could be made with 
software frameworks such as Siconos~\cite{acary2019introduction} and Project Chrono~\cite{projectChronoWebSite}, which are established for non-smooth problems. However, their compiled back-ends complicate the attribution of runtime differences relative to our Python implementation and thus are not used here.

Adaptive time-stepping methods dynamically adjust the time step to balance computational efficiency and accuracy. They estimate local errors by comparing solutions obtained from a single large time step with those from two smaller steps, as discussed in~\cite{hairer1993solving}. This approach is beneficial for maintaining accuracy while avoiding unnecessary computational effort in systems with rapidly changing dynamics.

By first performing a Taylor series expansion of the integration scheme, the local error $e_k$ can be approximated by:
\begin{equation} 
e_k = \frac{w_{1/2} - w_1}{2^p - 1}, 
\end{equation}
where $w_1$ is the solution after one large time-step, $w_{1/2}$ is the solution after two smaller steps, which are half of the large time-step, and $p$ is the order of the integration method. The relative error measure can then be expressed as~\cite{hairer1993solving,acary2008numerical}
\begin{equation} \label{eq: local error}
E_r = \left| e_k \cdot \text{tol}^{-1} \right|,
\end{equation}
where tol is a specified tolerance for error control. Thus, the optimal time-step is given by~\cite{hairer1993solving,press2007numerical,acary2008numerical}:
\begin{equation}\label{eq: optimal step}
    h_{\text{opt}}= h \left( \frac{1}{E_r}\right)^q,
\end{equation}
where $q=\frac{1}{p+1}$ and $h$ is the previous time-step size, and is equivalent to a P controller adaptive time stepper~\cite{soderlind2002automatic}.

To ensure stability and accuracy, the optimal time step is adjusted using a heuristic rule that prevents the time step increment from experiencing extreme variations. The empirical-based/heuristic update for the time-step, $ h_{\text{new}}$, is given by~\cite{hairer1993solving,acary2009toward}:
{\footnotesize
\begin{equation}\label{eq:steps}
   h_{\text{new}} = 
\begin{cases}
    \min\left( h \cdot E_r^{-q}, h_{\text{max}} \right), & \text{if successful and }  E_r > 1, \\
    \max\left( h \cdot \min\left( c_{\text{up}}, E_r^{-q} \right),h_{\text{min}}, t_{\text{max}} - t \right), & \text{if successful and } E_r \leq 1, \\
    h \cdot c_{\text{down}}, & \text{if not successful and } h > h_{\text{min}}, \\
    \text{Abort}, & \text{otherwise}.
\end{cases} 
\end{equation}}
Here, $h_{\text{min}}$ and $h_{\text{max}}$ are the minimum and maximum allowed time step, respectively, $c_{\text{up}}$ and $c_{\text{down}}$ are scaling factors that moderate the time-step size ratio $\frac{ h_{\text{new}}}{h}$, and are empirically determined.

While elementary heuristics like the one above are widely used, they can lead to oscillatory step-size sequences. A general framework based on a PI-controller-like approach to provide smoother step-size evolution is given by~\cite{soderlind2002automatic,soderlind2006adaptive,arevalo2020software}
\begin{equation}\label{eq:pi}
h_{\text{new}} = h \left(E_r^{-q/\gamma_1}  E_{r,n-1}^{q/\gamma_2} \frac{h_{n-1}}{h_{n-2}}^{-\kappa}\right),
\end{equation}
where $\gamma_1$, $\gamma_2$, and $\kappa$ are gains (or filter coefficients) chosen to achieve specific stability and smoothness properties. In contrast to the previous heuristic adaptive time stepping method (see~\cref{eq:steps}), rejection of a time step does not occur when $E_r>1$. Instead, the time step is rejected and reduced when the term $E_r^{-q/\gamma_1}  E_{r,n-1}^{q/\gamma_2} \frac{h_{n-1}}{h_{n-2}}^{-\kappa}< c_1$ and if $E_r^{-q/\gamma_1}  E_{r,n-1}^{q/\gamma_2} \frac{h_{n-1}}{h_{n-2}}^{-\kappa}> c_2$ then the step increase is capped by $c_2$~\cite{arevalo2020software}. The values of $c_1$ and $c_2$ are given in~\cite{arevalo2020software} and taken the same in the next paragraphs.

While the above approaches can be effective for smooth integration problems, the processes studied in this paper are generally nonsmooth. Nonsmooth dynamics involve abrupt changes in the system's behavior, making it challenging for traditional adaptive methods to maintain accuracy without high computational costs. Although conventional heuristic-based methods are not well-suited to nonsmooth dynamics, we use the conventional heuristic-based adaptive time-stepping method~\cref{eq:steps} as a baseline for comparison with the Reinforcement Learning-based adaptive time-stepping policy. For completeness, we also confirm the RL has better performance compared to the PI control approach mentioned above. This comparison enables us to assess the benefits of the RL-based adaptive time stepper for nonsmooth dynamics.

\section{Reinforcement Learning}
\label{RL: description}

Reinforcement Learning (RL) is a branch of machine learning where an agent learns to make decisions by interacting with an environment to maximize a cumulative reward~\cite{sutton2018reinforcement,stathas_rl,b:https://doi.org/10.1002/nag.3923}. In the context of adaptive time-stepping, the RL agent aims to learn a policy that selects optimal time steps to balance computational speed (or run time) and solution accuracy.

\subsection{Algorithm} 
We utilize the Truncated Quantile Critics (TQC) algorithm~\cite{kuznetsov2020controlling}, implemented within Stable-Baselines3~\cite{raffin2021stable}, which excels in continuous control tasks where actions are chosen from a continuous range rather than a set of discrete options. TQC belongs to the family of actor-critic methods, where an actor network selects actions and one or more critic networks evaluate how good those actions are by estimating expected future rewards. Unlike deterministic methods such as TD3~\cite{fujimoto2018addressing}, which use two critics to predict a single action-value and reduce overestimation by taking their minimum, TQC is distributional. It uses noise on target actions (target-policy smoothing) and careful update scheduling (delayed policy updates). Multiple quantile critics approximate the full distribution of returns, and the highest (most optimistic) quantiles are truncated, yielding a more conservative and stable value signal. In adaptive time-stepping with nonsmooth dynamics, this typically reduces overly aggressive step choices and produces steadier learning. A high-level overview of the TQC training loop is provided in~\cref{subsec: general_training}. To contextualize TQC’s performance, we additionally evaluate TD3 using the Stable-Baselines3 implementation~\cite{raffin2021stable}. TD3 provides a strong deterministic baseline for continuous control, allowing us to isolate the effect of distributional critics and quantile truncation on adaptive time-step selection.

\subsection{Architecture}

For the TQC , we use an actor-network with two hidden layers, each with 64 nodes, and a critic network with three hidden layers with 64 nodes. The larger number of hidden layers in the critic network is thought to enhance the agent's ability to accurately evaluate actions in environments with nonsmooth dynamics. Furthermore, following the original paper~\cite{kuznetsov2020controlling}, we employed five critic networks to provide a more robust estimate of the return distribution. Finally, we employed generalized state-dependent exploration to ensure a balanced exploration of the action space, thereby exploiting learned patterns. All hyperparameters other than those discussed above take default values according to StableBaselines3~\cite{raffin2021stable}. A summary of the hyperparameters used for the TQC algorithm is listed in~\cref{tab:tqc_hyperparameters}.

For the TD3, we use the default hyperparameters from the original paper~\cite{fujimoto2018addressing} and apply Gaussian noise to the action space during training, with a standard deviation of 0.1 to encourage action exploration. We use this simple implementation because we are interested in assessing out-of-the-box performance rather than fine-tuning either RL algorithm. For computational parity, we only adjust the network widths to match our TQC setup.

Note that the hyperparameters of the RL schemes used here (TQC and TD3) were the default ones. A comprehensive hyperparameter study of the RL schemes and their effects on performance, given their inherent capabilities, is beyond the scope of the present study. The only new hyperparameters introduced here are the parameter $\alpha$, for which a sensitivity analysis was performed (see~\cref{sec: example}).

\begin{table}[htbp]
\centering
\caption{TQC hyperparameters used in the Reinforcement Learning environment.}
\label{tab:tqc_hyperparameters}
\begin{tabular}{ll}
\hline
\textbf{Hyperparameter} & \textbf{Value} \\
\hline
Actor-network size & (64,64) \\
Critic network size & (64,64,64) \\
Number of critic networks & 5 \\
Learning rate & $3 \times 10^{-4}$ \\
Buffer size & $10^6$ \\
Batch size & 256 \\
Discount factor ($\gamma$) & 0.99 \\
Polyak coefficient ($\tau$) & 0.005 \\
Generalized state-dependent exploration & True \\ 
\hline
\end{tabular}
\end{table}

\subsection{Reward function}\label{sec: reward function}

The reward function is designed to explicitly encode the goals of adaptive time stepping: take steps as large as possible (to finish the simulation in fewer steps), keep each accepted step as cheap to compute as possible, and encourage accuracy. We presume the reward function has the form below
\begin{equation} \label{eq: reward}
\begin{cases}
       R= \left( \frac{h - h_{\text{min}}}{h_{\text{max}} - h_{\text{min}}} \right) \left( \frac{t_{\text{max}}^{\text{rt}} - t^{\text{rt}}}{t_{\text{max}}^{\text{rt}} - t_{\text{min}}^{\text{rt}}} \right) e^{-\alpha E_r}  & \text{if successful }, \\
          R= -\left( \frac{h - h_{\text{min}}}{h_{\text{max}} - h_{\text{min}}} \right) & \text{if not successful }.
\end{cases}
\end{equation}

The first term rewards choosing a large time step. Specifically, $\frac{h-h_{\text{min}}}{h_{\text{max}}-h_{\text{min}}}$ maps the chosen step size $h$ to the interval [0,1], so that values closer to 1 correspond to more aggressive stepping within the allowed range $[h_{\text{min}},h_{\text{max}}]$. On its own, however, always pushing $h$ upward can force the nonlinear solver to take many iterations, thereby increasing the wall-clock cost. To counter this, the second term $\frac{t_{\text{max}}^{\text{rt}} - t^{\text{rt}}}{t_{\text{max}}^{\text{rt}} - t_{\text{min}}^{\text{rt}}}$ rewards low per-step runtime. We measure the wall-clock time $t^{\text{rt}}$ spent solving the current step and normalize it, which again lies in [0,1] and is larger when the step was cheap to compute. In practice, $t_{\text{min}}^{\text{rt}}$ can be estimated from a ``cheap'' converged step (e.g., one iteration), and $t_{\text{max}}^{\text{rt}}$ can be approximated from the observed worst-case iteration count and the solver's per-iteration cost; we find that these bounds are sufficient for stable training.

The third term encourages accurate local solutions. We use the dimensionless relative error $E_r\geq 0$ (see~\cref{eq: local error}) and include a factor $e^{-\alpha E_r}$ with $\alpha>0$. This term is close to 1 when the estimated local error is small and decays toward zero as the estimated error grows. Therefore, if the error term goes to 0, the reward is dominated by the relative step size and runtime. The scalar $\alpha$ is the only user-chosen parameter. A larger $\alpha$ increases the penalty for errors and therefore pushes the policy towards smaller, safer steps. Smaller $\alpha$ tolerates larger local error in exchange for speed. Note that other decreasing functions, instead of the exponential, with respect to $E_r$ span from 0 to 1 as well. The exponential function was sufficiently regular for the purposes of this study.

Finally, if the step fails to converge in the nonlinear solver, we assign a negative reward. The negative reward penalizes failed ambitious steps more than failed cautious steps. This teaches the policy to back off after a divergence event rather than repeatedly proposing unrealistically large steps. 

The reward function shares some aspects with previous RL-based adaptive time-steppers. More specifically, the reward is structured such that it encourages the largest possible time-step~\cite{dellnitz2023efficient}, considers computational time~\cite{dong2022adaptive,xu2023adaptive}, and penalizes the reward given a measure of error~\cite{dellnitz2023efficient}. Notice, however, that the negative reward term of~\cref{eq: reward} is not considered in~\cite{dellnitz2023efficient,dong2022adaptive,xu2023adaptive}, but in the case of nonsmooth problems, where non-convergence is more unlikely, it is essential. More generally, it is worth emphasizing that the definition of the reward function encodes the main aspects for accelerating the time integration of nonsmooth problems, as it will be showcased in~\cref{sect: RL perfomance}.

Within this reward, the only free hyperparameter is the scalar $\alpha$, which controls the trade-off between accuracy and speed. A sensitivity analysis covering three orders of magnitude of $\alpha$ is presented in~\cref{sec: example}. As a rule of thumb, users should use $\alpha=\frac{1}{q}$, where $q=\frac{1}{p+1}$ to approach the optimal time step (optimal in the sense of Appendix~\ref{sec: heuristic} and to decrease $\alpha$ below $\frac{1}{q}$ to prioritize speed over accuracy or conversely increase $\alpha$ above $\frac{1}{q}$ for higher accuracy. However, the above rule of thumb might not always hold for nonsmooth problems, as those studied in this paper. Our experience suggests $\alpha\in[0.5, 16]$ covers the most practical use case explored here, with the ideal value often around $\alpha\in[1,8]$ (see~\cref{sec: example}), depending on the particular dynamical system.

\subsection{Observational space}

A key advantage of the RL-based adaptive policy is its ability to learn intricate nonlinear relationships between the environment's observations and the actions taken, which conventional heuristic time-stepping methods have not been developed to do. To benefit from this, the observation space of the proposed RL approach is expanded to include physical quantities and quantities related to the nonlinear solver.

Previous works have expanded the observational space by providing the neural network with several numerical performance measures, such as the number of nonlinear iterations, the error of the nonlinear iterations, and the relative error of time-steps~\cite{dong2022adaptive,xu2023adaptive}. To provide a notation of state, the right-hand side of the considered ODE has been passed as an observation in the case of scalar equations~\cite{xu2023adaptive,dellnitz2023efficient}, or a latent space representation of the full state generated by an autoencoder~\cite{han2021artificial} was passed as a part of the observational space.

We take a similar approach here by providing the neural network with the same numerical performance measures and including the normalized time step $\hat{h}= \frac{h-h_{\text{min}}}{h_{\text{max}}-h_{\text{min}}}$. However, as we are concerned with a general number of different state variables which in turn may vary in space, we opt to provide simply the spatial average of each state $\bar{w}$ and the spatial average of the internal energy function $\bar{U}$ as opposed to providing either the entire state vector or compounding the complexity of the problem by training an autoencoder to reduce the full state to a latent variable representation. By including this term as an observation, we incorporate a measure of the system's energetic state directly into the neural network.

To this end, the observational space is defined as, 
\begin{align*}
&\mathcal{O} = \{ \hat{U}(w), \hat{w}, \frac{{E}_{\text{iter}}}{E_c}, \frac{1}{1+{E}_r}, \hat{h}, \lambda\},\\
&\hat{U}(w)=\frac{\bar{U}(w)}{U_c} \;\in\; [0,1+\epsilon],\\
& \hat{w}= \frac{\bar{w}}{w_{\text{max}}} \;\in\; [0,1+\epsilon],\\
&\hat{h} = \mathcal{I}\frac{h-h_{\text{min}}}{h_{\text{max}}-h_{\text{min}}},
\end{align*}
where $\widehat{U}(w)$ is the normalized energy function, $U_c$ is defined as some critical value which is discussed in~\cref{sect: RL perfomance}, $\hat{w}$ is the normalized representation of the scalar representation of a given state, $\lambda$ is a boolean function for whether convergence was achieved, $\mathcal{I}$ is a function that returns -1 if the step did not converge and 1 if it did, and $\epsilon$ is some small number to be discussed below, ${E}_{\text{iter}}$ is the summation of the error returned from the iterative nonlinear solver, which is not equivalent to $E_r$. Note that the normalization by $E_c$ was determined by experimentation to ensure it did not exceed a value of 1.

Typically, the state's exact maximum value is unknown before the simulation. Similarly, the so-called critical value of the internal energy function $U_c$ may vary or not be clearly defined. However, these parameters can typically be estimated for a particular physical problem, thereby providing a normalization that yields a maximum normalized value of approximately 1. Thus, we assume the observational space may be a small value $\epsilon$ beyond 1. More robust techniques could be devised, e.g., an environment that adaptively updates the normalization, but such implementations were unnecessary for the scenarios considered within this work. 

\subsection{General training procedure for TQC}
\label{subsec: general_training}
\begin{figure*}[t]
\centering
\includegraphics[width=0.8\textwidth,height=0.6\textheight,keepaspectratio]{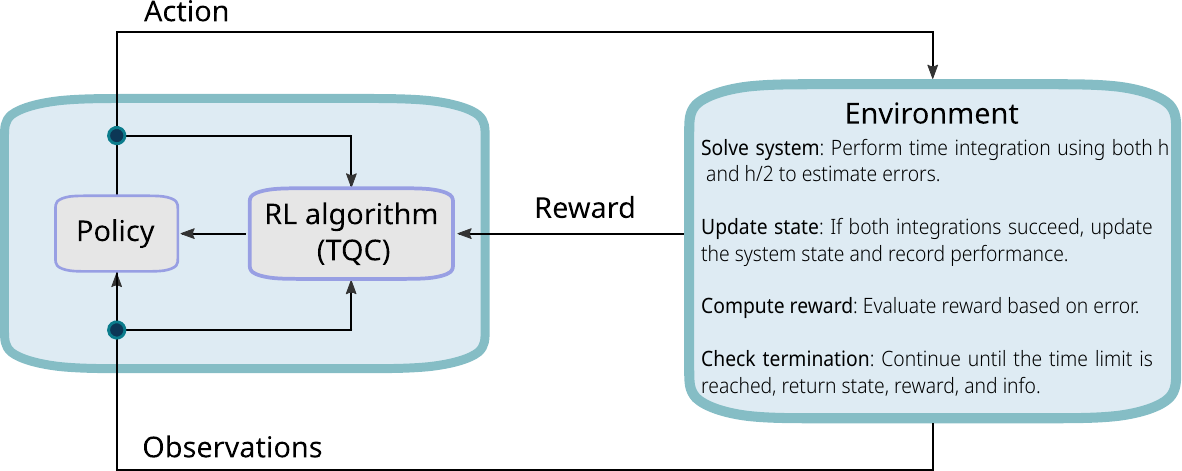}
\caption{Illustration of the logic applied within the Reinforcement Learning algorithm framework.}
\label{fig: 1}
\end{figure*}

The overall workflow of the algorithmic structure is illustrated in Figure~\ref{fig: 1}, which shows the interplay between the policy (actor networks), the TQC algorithm, and the environment. For each attempted time step, the environment receives the current action, i.e., a proposed time step $h$, and advances with both a single large time step, $t^{n+1}=t^{n}+h$, and a substepping method with two smaller time steps of size $\frac{h}{2}$. These two integrations enable the environment to estimate the local error by~\cref{eq: local error}, which is required for the reward. The environment updates the observational space, except for the observed quantities $\widehat{U}$ and $\hat{w}$ if neither integrator converges. 

Following an attempted time step, the TQC algorithm generates a ``transition" comprising the prior observations, the action taken, the resulting reward, and the updated observations, which are then stored in a replay buffer. Once a set number of transitions is collected, the TQC algorithm begins training at a specified interval. In our work, training occurs after each attempted time step, similar to previous studies using RL-based adaptive time integration~\cite {dong2022adaptive,xu2023adaptive}.

The training procedure in the TQC algorithm, as implemented by Stablebaselines3~\cite{raffin2021stable} based on~\cite{kuznetsov2020controlling}, is as follows. During training, the algorithm first calculates the target value by the sum of the immediate reward and the discounted estimate of the future returns. More specifically, copies of the critic networks, often called ``target networks'', provide these estimates of future returns to avoid rapid swings in learning.
Next, the difference between the predictions of the main critic networks and the target value is minimized via gradient descent, so that the critics' updated parameters yield better predictions of the actual outcomes. To prevent abrupt shifts in the target estimates, the target networks' parameters are slowly updated by a small factor termed the soft update coefficient (or Polyak coefficient), $\tau=0.005$. The policy network's parameters are then updated via gradient-based optimization to select better time steps that yield higher returns for the main critics' newly updated distributions, thereby maximizing the agent's predicted long-term performance. For a more thorough discussion of the algorithm's nuances, readers are referred to the original work~\cite{kuznetsov2020controlling}.

Finally, we manually explored the number of training steps and found that 20,000 steps resulted in a plateau in reward across all scenarios. All TQC and TD3 networks were trained on a laptop with the following specifications: AMD Ryzen 9 8945HS processor, 32 GB RAM, and an NVIDIA GeForce RTX 4070 Laptop GPU. Each training scenario took at most ten minutes for the scenarios in~\cref{sect: RL perfomance}.

It should be noted that the environment selects a time step solely based on the observational space, i.e., we do not enforce any changes to the time step chosen by the network, unlike previous studies~\cite{xu2023adaptive}. However, this lack of constraint on the policy implies that if the TQC (or TD3) network is not adequately trained, it may continually attempt time steps that may not converge, resulting in a constant observational space. 

\section{RL adaptive time-stepping performance}
\label{sect: RL perfomance}
We evaluate the general performance of the RL-based adaptive time stepper in handling nonsmooth dynamics. To test the robustness and adaptability of the RL-based method, we simulate three progressively more complex, nonsmooth dynamical systems. First, we explore the simple scenario of a smooth dynamical system with a first-order sliding mode controller, which introduces nonsmooth behavior. Next, we simulate an electrical circuit with a resistor that is a set-valued function of the state, similar to~\cite{matsumoto1985double}. Finally, we explore a challenging fault-mechanics scenario characterized by spatiotemporal Coulomb frictional instabilities. It is worth emphasizing that all simulations employ Bathe's implicit integration method~\cite{bathe2012insight}, but other integration schemes could also be used.

Additionally, since the speed-accuracy trade-off parameter $\alpha$ (\cref{eq: reward}) is the only new hyperparameter introduced, we perform a parametric analysis by training independent TQC networks for $\alpha = \{0.5,1,2,4,8,16,32,128\}$~in \cref{sec: example}, where $\alpha$ spans two orders of magnitude.

The parameters used for the heuristic time-stepper, as shown in~\cref{tab: heuristic_param}, were determined through trial and error to achieve the best performance in all simulated scenarios. Additionally, the best PI controller parameters (i.e., $\gamma_1$, $\gamma_2$, and $\kappa$) were selected from the various recommendations in~\cite{arevalo2020software}, which performed best across the scenarios considered in this paper aside from the degenerate case that aligns with~\cref{eq:steps}. Note that despite Bathe's method having an integration order $p=2$ for smooth systems~\cite{bathe2012insight}, we opted for $p=1$ to account for the possibility that the numerical integration scheme experiences a reduction in order as observed in both nonsmooth systems~\cite{acary2008numerical} and stiff systems~\cite{soderlind2002automatic}. Additionally, a relatively small initial time step was taken. Given that the heuristic stepper reduces the time step until convergence, this choice of the initial time step resulted in faster integration. All case studies used the parameter set in~\cref {tab: heuristic_param} for the conventional adaptive integration, unless stated otherwise.

\begin{table}[htbp]
\centering
\caption{Heuristic adaptive time-stepper hyperparameters.}
\label{tab: heuristic_param}
\begin{tabular}{ll}
\hline
\textbf{Hyperparameter} & \textbf{Value} \\
\hline
Integrator order ($p$) & 1 \\
Minimum time-step ($h_{\text{min}}$) & 0.00001 [s] \\
Time-step increasing scaling factor ($h_{\text{up}}$) & 2. \\
Time-step decreasing scaling factor ($h_{\text{down}}$) & 0.6 \\
\hline
\end{tabular}
\end{table}

\begin{table}[htbp]
\centering
\caption{PI adaptive time-stepper hyperparameters.}
\label{tab: pi_param}
\begin{tabular}{ll}
\hline
\textbf{Hyperparameter} & \textbf{Value} \\
\hline
Integrator order ($p$) & 1 \\
Minimum time-step ($h_{\text{min}}$) & 0.00001 [s] \\
$\gamma_1$ & 1/6 \\
$\gamma_2$ & 1/6 \\
$\kappa$& 1/6  \\
$c_1$ & 1.8\\
$c_2$ & 2.2 \\
\hline
\end{tabular}
\end{table}
Python’s time package measured computational speed, and we averaged 10 simulations to reduce variability from background processes. Additionally, the code was given the highest priority possible on the computer to minimize further variance introduced by background processes.

\subsection{Case Study 1: First-order sliding mode controller}
Here, we showcase the simplest scenario of a first-order sliding mode controller. The dynamical system is given as
\begin{equation} \label{eq: sliding mode}
    \dot{w} =\phi(t) + F_r
\end{equation}
where $w$ is the sliding variable, $\phi(t) = D \text{sin}(\omega t)$ is a bounded disturbance of amplitude $D$ and $\omega$ is the angular frequency, $F_r \in - k \text{sign}(w)$ is the sliding mode controller input. To ensure convergence to the sliding surface, $w\to0$, in finite time, it is required that $k>D$ as shown in~\cref{app0}. The constraint $F_r$ can be represented by the closed convex set
\begin{equation} \label{eq: sliding_cone}
    \mathcal{C}= \{(F_r,k) \in \mathbb{R} \times \mathbb{R}_{+}| ~~~|F_r| \leq k\},
\end{equation}
which enables this problem to be equivalently expressed as a variational inequality~\cite{acary2008numerical}, thereby satisfying the nonsmooth controller input.

\subsubsection{RL training for Sliding mode controller}
\label{sect: cs1}
The TQC network, with the trade-off parameter $\alpha=2$, uses a continuous action space for time step $h \in [0.00001, \frac{2\pi}{10\omega}]$ [s]. These bounds were chosen such that $h_{\text{min}}$ guarantees convergence, while the maximum time step, $h_{\text{max}}$, is ten times the bandwidth of the closed-loop system, in accordance with recommendations in control applications~\cite{astrom1997computer}. Although larger maximum time steps could be used, we chose a truncated action space for this example. 

For the observational space, we use the maximum state possible~\cref{eq: bound}, which is determined in~\cref{app0}, for the normalization of the instantaneous value of the state as well as for calculating the critical internal energy of the system $U_c$.

 For training, the dynamical system parameters are set as $k=2$ [1/s], $D=1.5$ [1/s], and $\omega=50$ [rad/s]. Finally, the training was performed with the initial condition $w(0)=0.1$.

\subsubsection{Result}

\begin{figure*}[t!]
\centering
\includegraphics[width=0.95\textwidth,height=0.45\textheight,keepaspectratio]{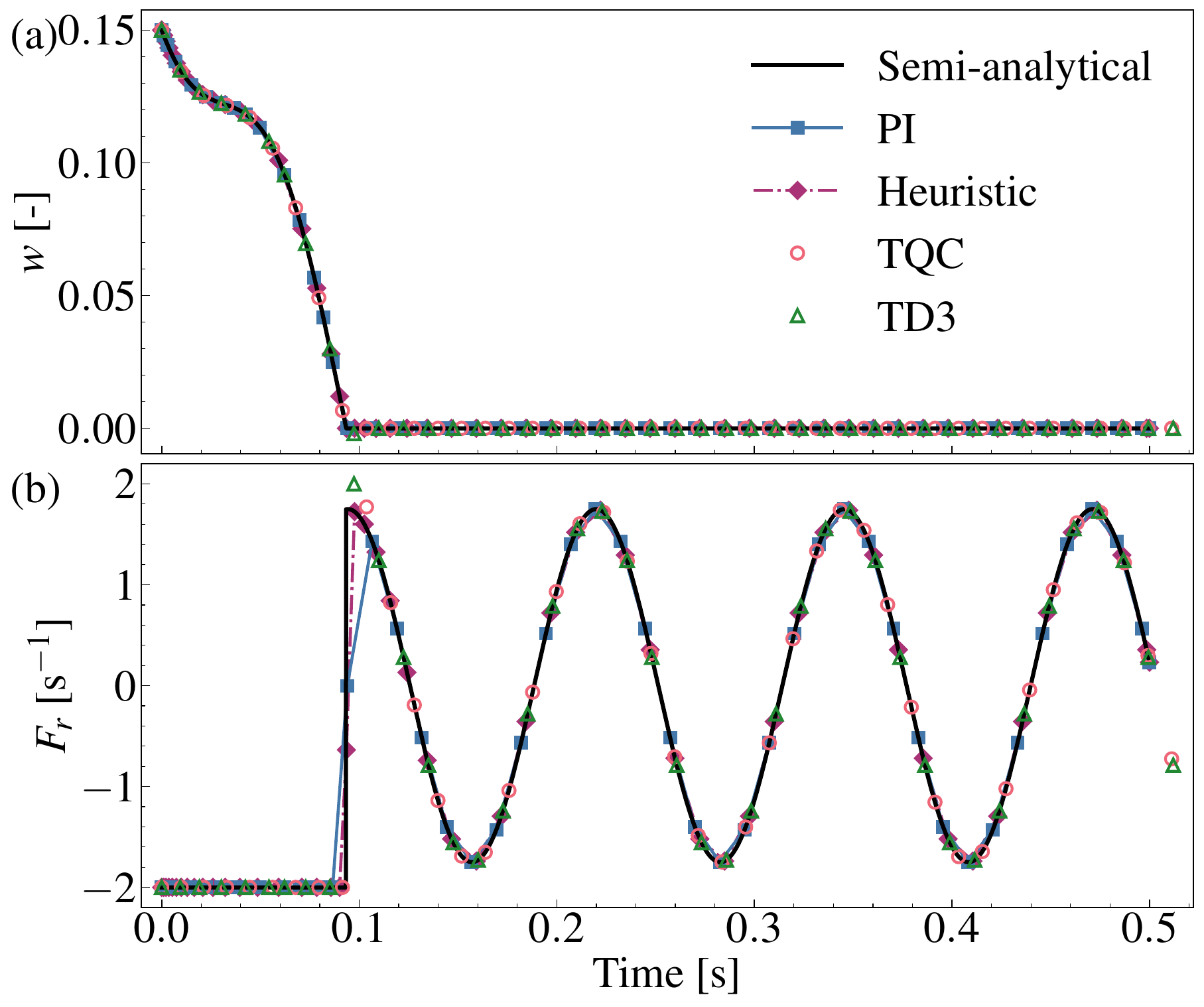}
\caption{(a) State $w$ against time and (b) controller input $F_r$ against time for a scenario that the TQC network was not trained on.}\label{fig: smc}
\end{figure*}

To evaluate the performance of the proposed RL-based adaptive time-stepper in this simple scenario, we set the maximum disturbance to $D=1.75<k$ and the initial condition to $w(0)=0.15$. For direct comparison, the maximum time step for the heuristic-based adaptive integrator is set to $\frac{2\pi}{10\omega}$.

~\cref{fig: smc} shows the results of the RL-based adaptive time integrator (TQC and TD3) compared to those of the heuristic adaptive integration, PI stepper, and a piecewise analytical solution. Notably, the RL-based adaptive integrator achieves approximately a 2× speedup over the heuristic-based method without sacrificing accuracy. In particular, the normalized runtimes of the PI, heuristic, TQC, and TD3, relative to the slowest, are 1.0, 0.90, 0.44, and 0.43, respectively. The corresponding L$_2$ norm errors with respect to the semi-analytical solution are $1.9\cdot10^{-4}$, $1.7\cdot10^{-4}$, $2.9\cdot10^{-4}$, and $2.7\cdot10^{-4}$ for the PI, heuristic, TQC, and TD3 methods, respectively. Importantly, the RL methods maintain accuracy at the same order of magnitude as the traditional approaches while providing a twofold speedup.

\subsection{Case Study 2: Electrical circuit}
\label{sect: cs2}
The previous scenario illustrates improvements achieved through RL-based adaptive integration in a simple setting. We now consider a more complex and realistic scenario inspired by Chua's circuit~\cite{matsumoto1985double}. The following dynamical system describes the electrical circuit:
\begin{equation}\label{eq: ec}
\begin{bmatrix}
\dot{V_1} \\
\dot{V_2} \\
\dot{i}
\end{bmatrix}
=
\begin{bmatrix}
- \frac{1}{C_1R} &  \frac{1}{C_1R} & 0 \\
 \frac{1}{C_2R} &  -\frac{1}{C_2R} & \frac{1}{C_2} \\
0 & -\frac{1}{L} & 0
\end{bmatrix}
\begin{bmatrix}
V_1 \\
V_2 \\
i
\end{bmatrix}
+
\begin{bmatrix}
 F_r(V_1) \\
0 \\
0
\end{bmatrix}
\end{equation}
where $V_1$ and $V_2$ denote voltages, $i$ is the inductor current, $C_1$ and $C_2$ are capacitances, $R$ is a resistance, and $F_r(V_1)$ is a diode voltage rate. Although in Chua's circuit, this is given by a piecewise linear function~\cite{matsumoto1985double}, we opt instead for $F_r\in-\frac{m}{C_1}\text{sign}(V_1)$ to emulate an ideal Zenner diode~\cite{acary2010nonsmooth}, which can be formulated in an equivalent form to~\cref{eq: sliding_cone}. 

\subsubsection{RL training for Electrical Circuit}
The TQC network with the trade-off parameter $\alpha=2$ a uses continuous action space $h \in [0.00001, \frac{t_\text{{max}}}{10}]$ [s]. The bounds were chosen such that $h_{\text{min}}$ would always converge and $h_{\text{max}}$ would yield a minimum of 10 time steps. We use each state $w$ and the system's internal energy as components of the observational space. Since in~\cref{app: ec} we show that the internal energy decreases asymptotically to zero, we normalize the energy~\cref{eq: ec_ie} by the internal energy at the initial condition $U_c=U_0$. We derive the normalization factors from this initial internal energy to normalize the individual states, ensuring that all normalized states $w\in[0,1]$.

The parameters of the dynamical system for training are set to be $C_1=\frac{1}{9}$ [F], $C_2=1$ [F], $R=\frac{1}{0.7}~[\Omega]$, $L=\frac{1}{7}$ [H] and the initial conditions $V_1= 0.15264$ [V], $V_2$ = -0.02281 [V], and $i$ = 0.38127 [A], which are chosen to match~\cite{matsumoto1985double}.

\subsubsection{Results}

\begin{figure*}[t]
\centering
\includegraphics[width=0.95\textwidth,height=0.7\textheight,keepaspectratio]{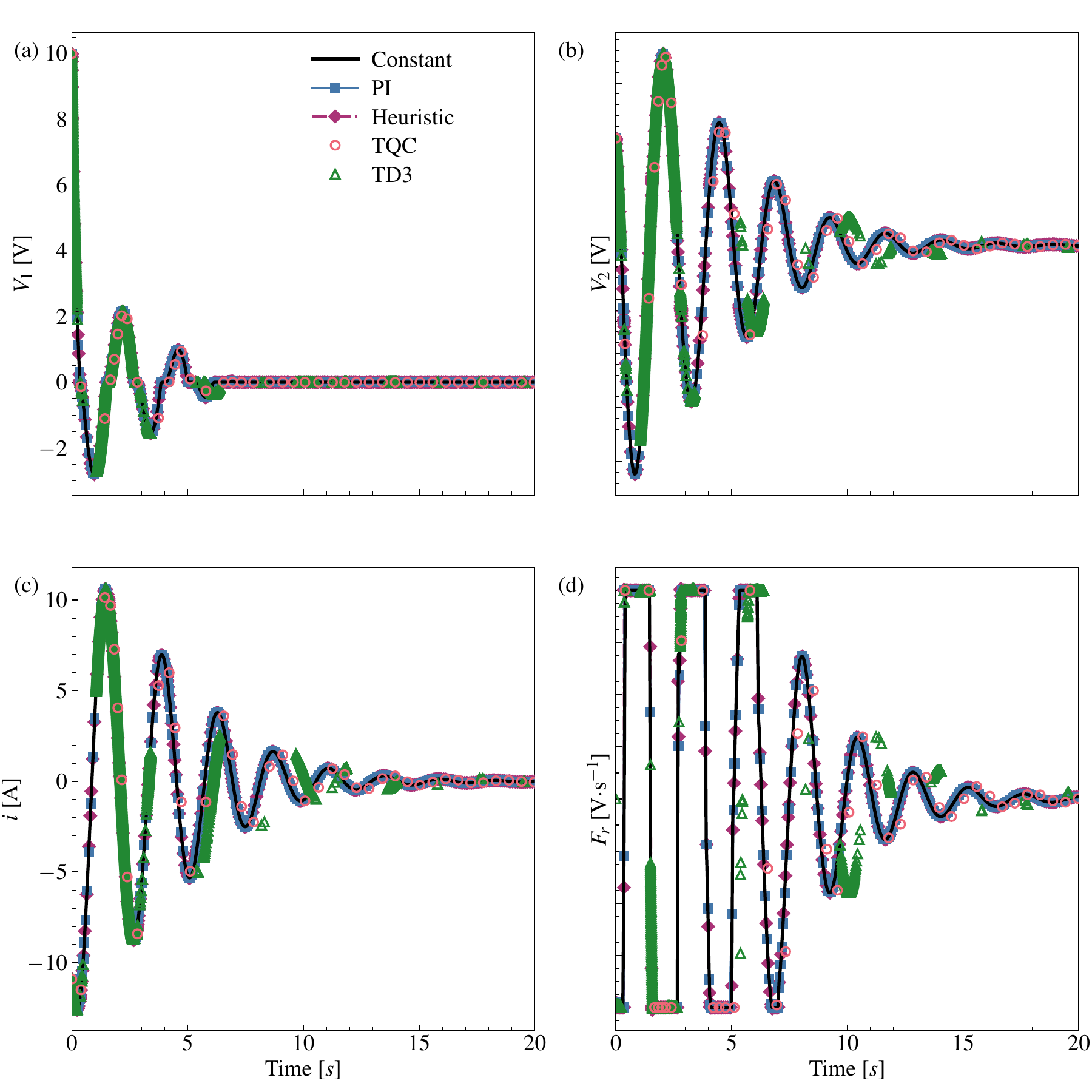}
\caption{(a-c) States $V_1$, $V_2$, $i$, and $F_r$ against time for constant, PI, heuristic, and RL-based integrators (TQC and TD3). Note that this evaluation is a scenario that the TQC network was not trained on.}\label{fig: ec}
\end{figure*}

To evaluate the TQC network in this simple scenario, we change the initial conditions to  $V_1$ =9.998048 [V], $V_2$ =1.980972[V], and $i$ = -10.908448 [A], which is also taken from~\cite{matsumoto1985double}. Critically, the normalization of constants for the internal energy and state is updated, provided this new initial condition .~\cref {fig: ec} shows the results from the TQC and TD3 adaptive time integrators along with the heuristic-based, PI integrator and a constant time stepper with $h=0.001$ [s].~\cref{fig: ec} shows the three states and $F_r$ against time. Naturally, this set-valued function gives rise to nonsmooth behaviour in $V_1$. 

TQC attains essentially the same accuracy as the PI controller and TD3 while being roughly two orders of magnitude faster in wall-clock time, because PI and TD3 both produce very fine time discretizations. Moreover, relative to the purely heuristic adaptive scheme that enforces $E_r\leq1$ at every step, TQC is over three orders of magnitude faster. In particular, the runtimes of the PI, heuristic, TQC, and TD3, normalized to the slowest, are 0.10, 0.13, 0.01, and 1.0, respectively. Thus, we find that the TD3 is by far the slowest. Impressively, the TQC is 10$\times$ faster than the PI and the heuristic. As all state variables decay to zero at steady state, we quantify accuracy using the relative time-averaged L$_2$ error in time with respect to the constant-step reference solution. The corresponding values are 0.38\%, 0.44\%, 19.6\%, and 27.2\% for the PI, heuristic, TQC, and TD3 methods, respectively. The improved calculation time of TQC is attributed to the drastic reduction in the number of evaluations over time, as seen in~\cref {fig: ec}. This indicates that the improvement is significant for nonsmooth dynamical systems with a small number of degrees of freedom.

\subsection{Case Study 3: Strike-slip fault}
\label{sec: example}
\begin{figure}[t]

\centering
\includegraphics[width=0.8\textwidth,height=0.8\textheight,keepaspectratio]{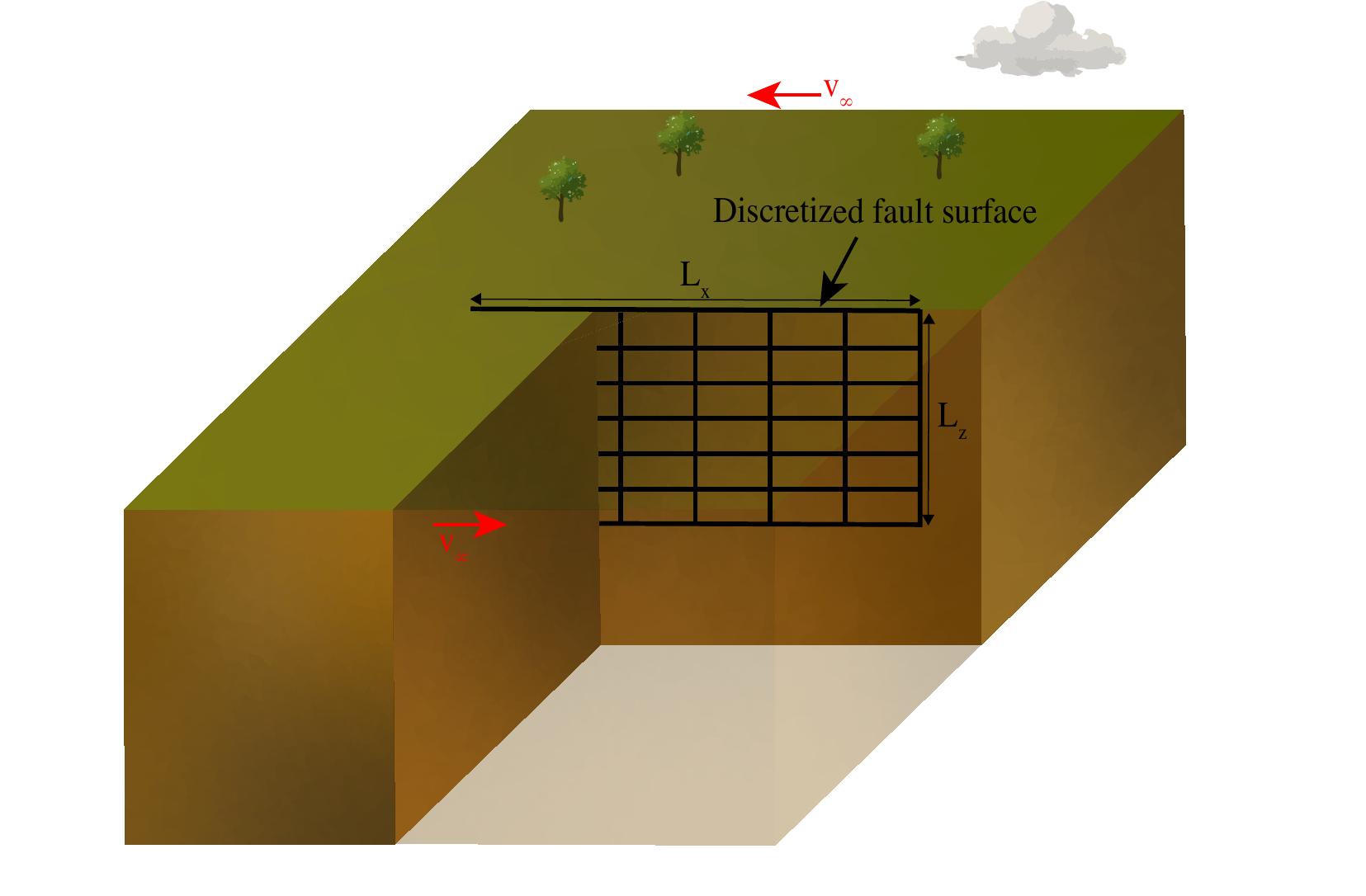}
\caption{Illustration of a strike-slip fault of dimensions $L_x$ and $L_z$ with a far-field velocity $v_\infty$.}\label{fig: fault}
\end{figure}

Finally, we present the dynamics of an isolated strike-slip fault configuration illustrated in~\cref{fig: fault}. This is a quite challenging example of a nonsmooth system as it has many more degrees of freedom than the previous examples. The fault is just beneath the surface, covering a rectangular area of $L_x \times L_z$ in the $x$- and $z$-directions, respectively. We also assume that the fault is adequately oriented in the tectonic stress regime for a slip to occur. For the spatial discretization of the fault area, we use a regular mesh with $N_x$ elements along the $x$-axis and $N_z$ elements along the $z$-axis. Therefore, the total number of elements covering the fault area is $N = N_x \times N_z$ with each element's size being $D_x \times D_z$, where $D_x = \frac{N_x}{L_x}$ and $D_z = \frac{N_z}{L_z}$. 

This system shares the same formulation as~\cref{eq: dynamics} and is described by the linear momentum equation in state representation:
\begin{equation}
\begin{bmatrix}
M & 0 & 0\\
0 & I & 0\\
0 & 0 & I
\end{bmatrix}
\begin{bmatrix}
\dot{v} \\
\dot{\delta}\\
\dot{s}
\end{bmatrix}
=
\begin{bmatrix}
- H & K & 0 \\
I & 0 & 0 \\
0 & 0 & I 
\end{bmatrix}
\begin{bmatrix}
v \\
\delta \\
|v|
\end{bmatrix} \\
+
\begin{bmatrix}
H v_{\infty} - K v_{\infty} t + F_r(t, s, |v|) \\
0\\
0
\end{bmatrix}
\end{equation}
where $K\in  \mathbb{R}^{N \times N}$ is the elasticity matrix, $H \in  \mathbb{R}^{N \times N}$ is the viscosity matrix, $v_{\infty} \in \mathbb{R}^{N }$ is a far-field velocity that continuously strains the system, $\delta \in \mathbb{R}^{N}$ is the displacement of the fault, and $F_r \in \mathbb{R}^{N}$ is the friction force (or constraint). A lumped mass matrix  $M=\rho L_y D_xD_zI_N$ was used to represent the mobilized mass during abrupt slip, where $I_N$ is the identity matrix of size $N$ and $\rho$ is the density. Furthermore, the viscosity matrix $H= 2 \zeta M \sqrt{M^{-1}K}$ where $\zeta$ is the damping ratio, and the square root is defined for matrices. Finally, the elasticity matrix $K= G K_{con}$ where $G$ is the shear modulus and $K_{con}$ is the connectivity matrix as in~\cite{rice1993spatio}. For more details, we refer to~\cite{stefanou2022preventing,gutierrez2024passivity}. Furthermore, the constraint force given by Coulomb friction is typically defined by the closed convex set
\begin{equation}
    \mathcal{C} = \{(F_r,q) \in \mathbb{R} \times \mathbb{R}_{+}| ~~~|F_r| \leq q\},
\end{equation}
where $q=\mu(s,|v|) A \sigma_n^{'}$ represents a variable determined by the frictional coefficient $\mu(s,|v|)$, where $s$ is the time-integrated value of $|v|$ (or slip-rate) and normal force $A\sigma_n^{'}$ is given by the area and normal effective stress to the fault plane.

In this work, we chose a slip weakening law~\cite{kanamori2004physics}
\begin{equation}\label{eq: swl}
\mu = \mu_{d}+\Delta\mu e^{-s/d_{c}},
\end{equation}
where $\mu_{d}$ is the dynamic friction value, $\Delta \mu$ is the difference between static and dynamic friction, and $d_{c}$ is the critical sliding distance. Alternative friction laws could be used, such as in~\cite{stefanou2022preventing}, but this would not be expected to alter the conclusions of this paper.

\subsubsection{RL training and heuristic parameters specifics}
\label{sec: training_specifics_fault}
The complete set of material parameters for the strike-slip fault is presented in~\cref{tab:fault_material_param,tab:fault_friction_param}, where the maximum estimated velocity $v_{\text{max}}^{\text{est}}$ is determined in the same manner as in~\cite{tzortzopoulos2021controlling}. In contrast to~\cref{sect: cs1,sect: cs2}, we instead use the spatially averaged quantity of the internal energy and only the velocity $v$ state variable in the observational space. We deviate from previous case scenarios because the states $\delta$ and $s$ monotonically increase, and rates are entirely encoded with the state $v$. To normalize the observed average velocity, we use the peak-estimated velocity. Additionally, the internal energy at the point of frictional instability was used to normalize the observational average internal energy.

\begin{table}[htbp]
\centering
\caption{Material and Geometric Parameters for the Fault Used for Simulations.}
\label{tab:fault_material_param}
\begin{tabular}{ccccccc}
\toprule
$L_x = L_z$ & $G$ & $(\sigma'_n)_{\text{avg}}$ & $\rho$ & $\nu$ & $\zeta$ & $v_\infty$ \\
{[km]} & {[GPa]} & {[MPa]} & {[kg/m$^3$]} & {[-]} & {[-]} & {[cm/year]} \\
\midrule
3 & 30 & 22.5 & 2500 & 0.25 & 0.27 & 0 \\
\bottomrule
\end{tabular}
\end{table}

\begin{table}[htbp]
\centering
\caption{Friction Parameters and Predicted Earthquake Velocity.}
\label{tab:fault_friction_param}
\begin{tabular}{cccc}
\toprule
$\Delta \mu$ & $\mu_{\text{res}}$ & $d_c$ & $v_{\text{max}}^{\text{est}}$ \\
{[-]} & {[-]} & {[mm]} & {[m/s]} \\
\midrule
0.1 & 0.5 & 100 & 0.07 \\
\bottomrule
\end{tabular}
\end{table}

As in the previous case studies, we use a continuous action space $h \in [0.00001, \frac{t_{\text{max}}}{5}]$ [s]. The bounds were chosen so that $h_{\text{min}}$ would always converge and $h_{\text{max}}$ would require at least 5 time steps to complete the simulation. However, contrary to previous applications, we now require spatial discretization specifications. For training purposes only, the fault is discretized into a single element, i.e., $D_x=D_z=L_x=L_z$. Although a single element does not represent a finely discretized simulation, it provides two advantages for training the TQC and TD3 networks. First, it provides a relatively fast training time. Second, it enables exploration of how well the proposed RL-based adaptive policy generalizes to finer discretizations. However, during evaluation, the adaptive time stepping with TD3 failed to converge systematically, i.e., it repeatedly failed to advance the time integration across all scenarios. Given that the TD3 has performed poorly in these more complex scenarios, we forgo further discussion of it in the forthcoming sections. Hence, for simplicity, we refer to the RL-based method as the TQC-specific method.

\subsubsection{Computational speed}
\label{sec: comp_speed}

To evaluate general speed performance, we explore scenarios with two different fault sizes: 3 [km] by 3 [km] and 5 [km] by 5 [km]. For each fault, we consider mesh discretizations varying from 25, 50, 75, and 100 square elements. Furthermore, provide independent comparisons of the TQC algorithm first against the commonly applied heuristic law (see~\cref{eq:steps}), which is the more widely adopted adaptive time-stepping method. For further confirmation, we then compare the PI adaptive stepper (see~\cref{eq:pi}) but only for the 3 [km] by 3 [km]. Finally, we evaluate the attempted and accepted time-step increments each method used to fully understand the different routes it took through time, integrating a chosen scenario.

\begin{figure*}[!t]
\centering
\includegraphics[width=1.05\textwidth]{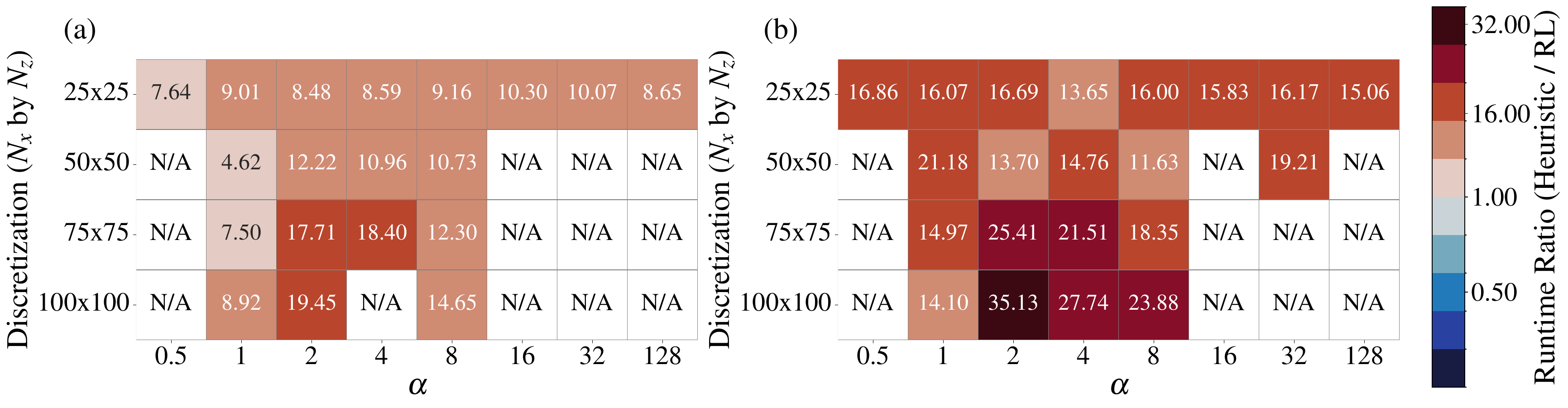}
\caption{Results demonstrating the ratio of the runtime of the heuristic-based method over the runtime of the RL-based method for (a) $L_x=L_z=3$ [km] fault and (b) $L_x=L_z=5$ [km] fault. The x-axis depicts the speed-accuracy trade-off parameter $\alpha$ (see~\cref{eq: reward}), and the y-axis depicts the fault discretization ($N_x$ by $N_z$). Note that the white boxes with the ``N/A'' text indicate that scenarios in the RL-based integrator did not converge.}\label{fig: speed_comp}
\end{figure*}

~\cref{fig: speed_comp} displays a tile plot for (a) fault configuration of $L_x=L_z= 3$ [km] and (b) fault configuration of $L_x=L_z= 5$ [km], where the y-axis represents different discretizations, the x-axis shows the values of the speed-accuracy trade-off parameter $\alpha$, and the color indicates the ratio of runtimes of the heuristic and RL time-steppers. Note that ``N/A'' in gray indicates scenarios where the RL-based policy did not converge. Here, non-convergence indicates that the RL-based policy repeatedly selected a time step that did not converge, and therefore, the simulation was terminated.

\cref{fig: speed_comp} (a) shows that aside from nonconvergent simulations, the RL-based policy is consistently faster than the heuristic-based policy by at least a factor of 4 and a maximum of 19.45. Notably, the relative speed advantage increases slightly as the mesh is refined. Regardless, the average speed improvement was roughly 11.02 times that of the heuristic for convergent scenarios. Interestingly, the speed advantage does not decrease monotonically with increasing $\alpha$, indicating a nontrivial dependence on this parameter and potentially slight variations in the training due to the stochastic nature of the TQC network.~\cref{fig: speed_comp} (b) shows that the RL-based algorithm also performs well on the 5 [km] by 5 [km] fault despite never being trained on that configuration. Again, across all convergent scenarios, the RL-based adaptive time stepper is at least 10 times faster and up to 35 times faster than the heuristic. Impressively, the average increase factor is 18 times. Similar to~\cref{fig: speed_comp} (a), we find similar trends for nonconvergence.

The nonconvergence occurs for $\alpha\leq0.5$ and $\alpha\geq16$ across all mesh discretizations except the coarsest, 25$\times$25. The inability of the $\alpha\leq0.5$ to converge is attributed to the potentially overly optimistic time steps. For $\alpha\geq 16$, the reward function becomes dominated by the error term, causing the gradient of the reward function with respect to $E_r$ to become extremely steep. Even tiny variations in estimated error cause massive swings in reward, making the policy optimization numerically unstable. The policy essentially loses the ability to distinguish between ``acceptably small" and ``slightly larger but still acceptable" errors, so that everything except near-zero error appears catastrophically bad. This can cause the policy to learn incoherent or oscillatory strategies, leading it to fail to settle on appropriate time step sizes.

\begin{figure*}[!t]
\centering
\includegraphics[width=0.65\textwidth]{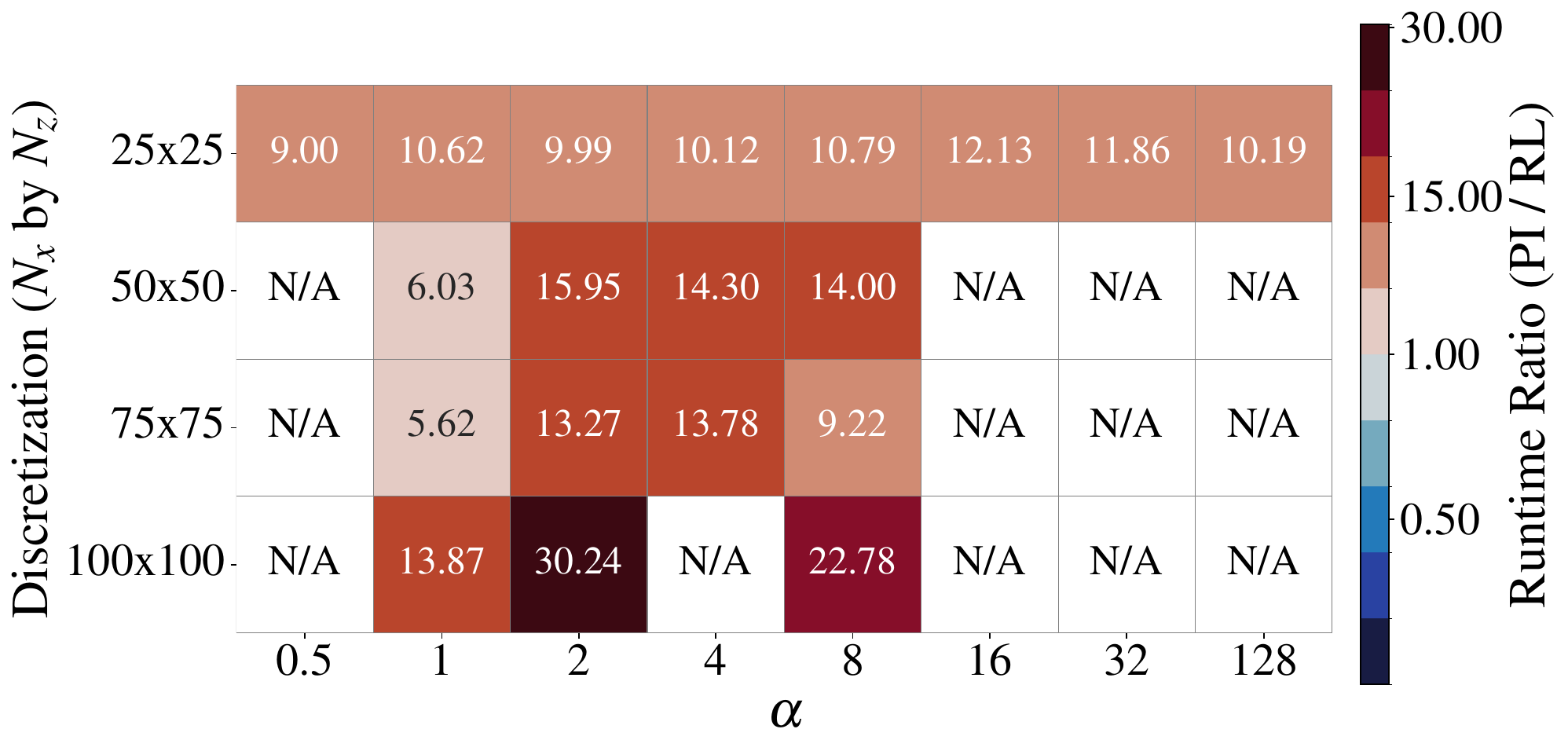}
\caption{Results demonstrating the ratio of the runtime of the PI-based method over the runtime of the RL-based method for (a) $L_x=L_z=3$ [km] fault and (b) $L_x=L_z=5$ [km] fault. The x-axis depicts the speed-accuracy trade-off parameter $\alpha$ (see~\cref{eq: reward}), and the y-axis depicts the fault discretization ($N_x$ by $N_z$). Note that the white boxes with the ``N/A'' text indicate that scenarios in the RL-based integrator did not converge.}\label{fig: speed_comp_pi}
\end{figure*}

Now \cref{fig: speed_comp_pi} (a) shows the same comparison but with respect to the PI adaptive time stepper. By observation of~\cref{fig: speed_comp_pi} (a), we find generally the same behaviour as in the previous figure. However, now the RL is at least 5 times faster and up to 30 times faster. The average ratio is 12.83, indicating that overall, the RL outperforms the PI adaptive stepper for wall-clock time. Note that the 100-by-100 discretization was run only once for the PI.


We find that the TQC consistently outperforms both the heuristic and PI stepper for convergent scenarios. Moreover, across all scenarios considered in this paper, the more advanced PI stepper generally performed at lower computational speeds. This is likely due to the nature of the past terms in~\cref{eq:pi}, which assume a degree of regularity in past errors that is not inherently clear for nonsmooth dynamics.

\begin{figure*}[ht]
\centering
\includegraphics[width=0.95\textwidth,height=0.7\textheight,keepaspectratio]{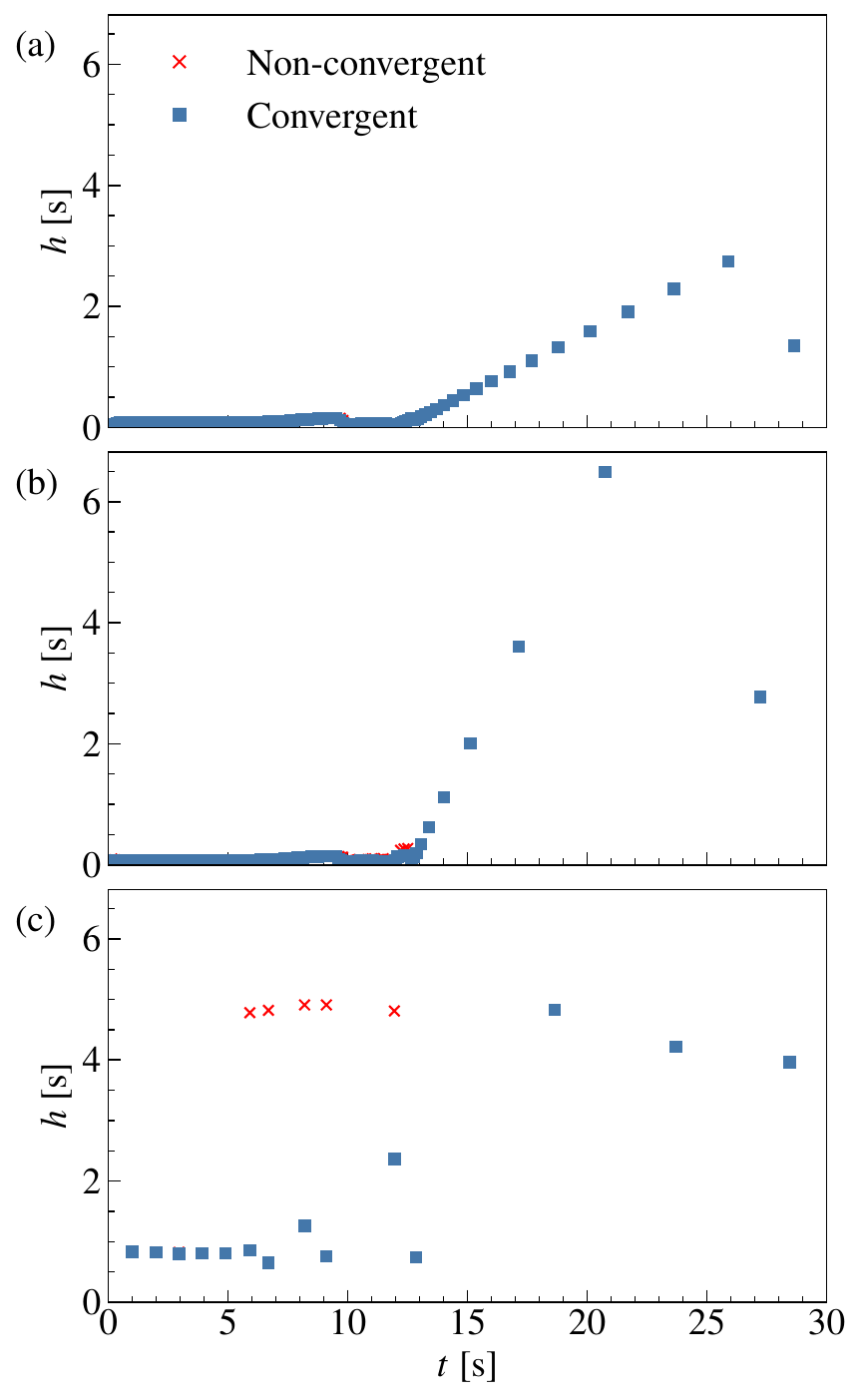}
\caption{Time step $h$ choice against time $t$ for the (a) PI time-stepper, (b) heuristic time-stepper, and (c) the RL-based approach.}\label{fig: h_evo}
\end{figure*}

To better understand why the RL-based simulation is generally faster, we examine the time steps selected by RL ($\alpha=2)$, heuristic, and PI approaches in the 3 [km] by 3 [km] scenario with a 50 by 50 mesh discretization, which also corresponds to the scenario in~\cref{fig: fault rl_accuracy}.~\cref{fig: h_evo} displays the time steps attempted by each approach over time, with red points representing non-convergent steps and blue points indicating convergent steps. The PI and heuristic take similar approaches. The PI approach shows that it initially takes very small steps until around $t=10$ [s]. At this point, increases in step-size attempts are met with step rejections, which could be due to a nonlinear solver's nonconvergence or to the adaptive scheme's innate rejection mechanism. Following this period, both policies gradually increase the step size during the sticking regime. However, the heuristic increases faster because it is not constrained by step-size increases as the PI stepper is. While the general trends mirror each other, this change in rate of increasing the step size is predominantly why~\cref{fig: speed_comp,fig: speed_comp_pi} indicates the heuristic is slightly faster.

In contrast, the RL-based approach learned to maintain nearly constant time steps (larger than those under heuristic/PI) at the initiation of slip events, followed by a sudden increase before plateauing during the sticking phase. Moreover, the steps that do not converge due to nonlinearities (rejected steps) are contrasted with a sudden large drop in the time step to ensure convergence. As a result, the RL takes many fewer time steps, since there is no strict rejection based on error, which likely leads to larger initial steps, and the sticking can increase instantly as there is no longer any dynamics. The apparent state-dependent strategy of the RL-based approach likely explains the improved speed in~\cref{fig: speed_comp} (b), as the RL-based policy leverages information of the average velocity state and internal energy to choose its time steps. 

\subsubsection{Accuracy}
\label{sec: accuracy}

The RL-based method, on average, provides faster simulations than the heuristic-based method and PI stepper, but a critical aspect is determining its accuracy. Here, accuracy is measured relative to the constant-time stepper with $h=0.05$ [s], as there is no closed-form solution for the scenarios explored in this paper. Note that upon evaluation, the difference in accuracy between the heuristic and PI stepper was less than 1e-5; thus, we present only the heuristic for comparison.

First, we illustrate the accuracy of RL-based and heuristic-based methods for a single scenario. The accuracy of the proposed RL-based time-stepping policy is first explored for a fault with a 50-by-50 element discretization and the speed-accuracy trade-off parameter $\alpha=2$. 

\begin{figure*}[t]
\centering
\includegraphics[width=0.95\textwidth,height=0.75\textheight,keepaspectratio]{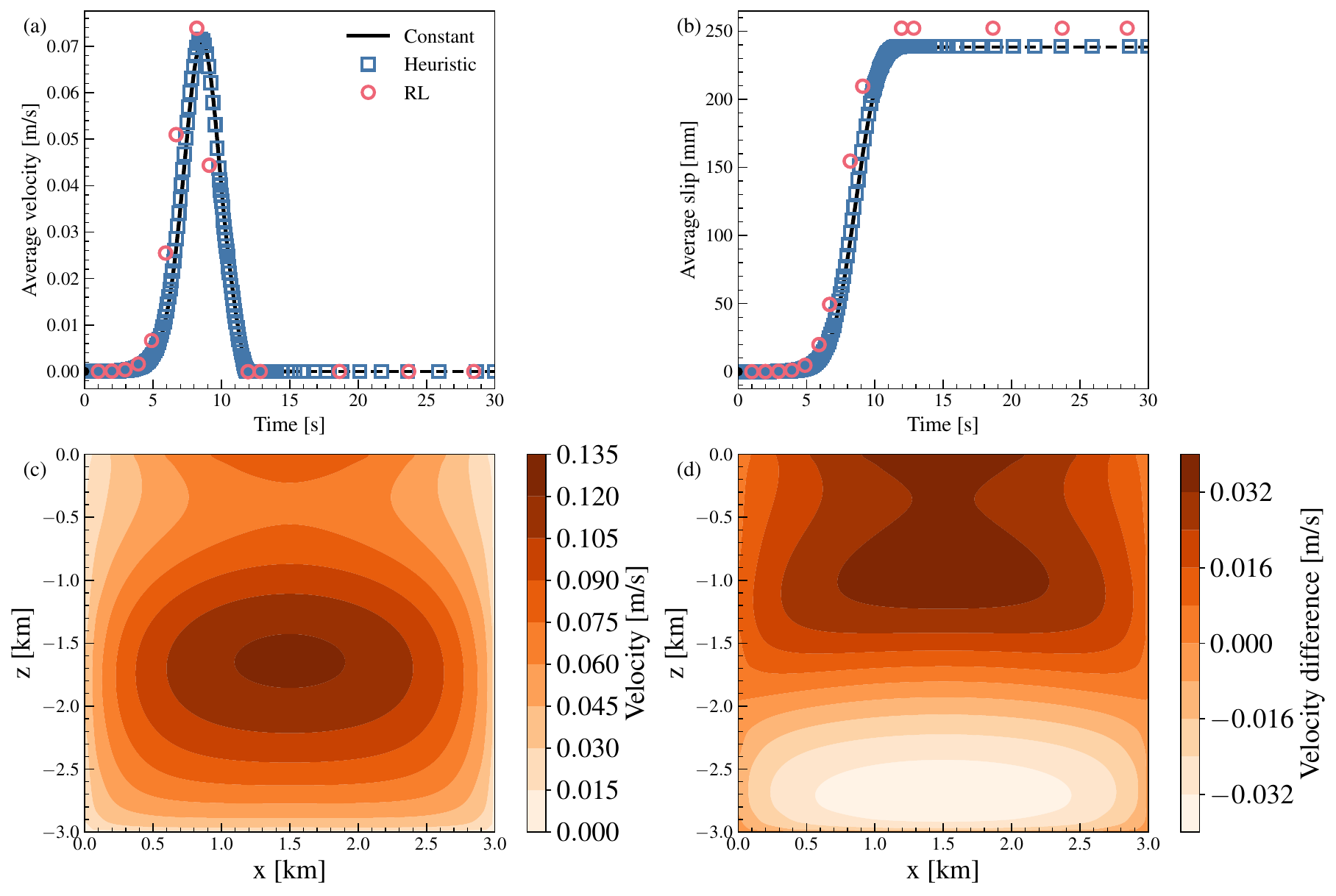}
\caption{Accuracy of the RL-based adaptive time-stepping approach for a frictional instability not in the RL training set. (a) Average velocity over time for three integration methods: constant time step ($h=0.05$ [s]), heuristic, and RL-based. (b) Average slip in time for the three integrators. (c) Velocity spatial distribution snapshot at $t=8.2$ [s] using the RL-based integrator. (d) Velocity difference spatial distribution snapshot between the RL-based and constant time step integration. Note that a dashed line shows the asymptotic average values of velocity and slip of the constant time-stepper.}\label{fig: fault rl_accuracy}
\end{figure*}

\cref{fig: fault rl_accuracy} (a) and (b) compare the RL-based method results of the average spatial velocity and average spatial slip over time against the heuristic method and constant time-stepping methods. By visual inspection, the RL-based method performs similarly to the heuristic method in terms of average spatial velocity, aside from a small time shift. A more comprehensive comparison to fully assess accuracy is the spatial velocity profile (c) displayed for the RL-based method at $t=8.2$~[s] and in~\cref{fig: fault rl_accuracy} (d) shows the velocity difference $v-v_{\text{true}} $, where $v_{\text{true}}$ is assumed to be given by the spatial profile of the constant time-stepper for $h=0.05$ [s]. We opt to display the difference because the relative error can be influenced by regions that have begun sticking. Notably, the maximum difference is roughly 0.03 [m/s], which is primarily attributed to the shifting we notice in~\cref{fig: fault rl_accuracy} (a). Regardless, the majority have relatively low differences, and as such, the RL-based adaptive time stepper captures not only the peak average velocity but the overall sliding behaviour. This preliminary evaluation suggests that both methods are relatively accurate, though a more comprehensive comparison is necessary to fully gauge accuracy across multiple scenarios.

\begin{figure*}[!t]
\centering
\includegraphics[width=1.05\textwidth]{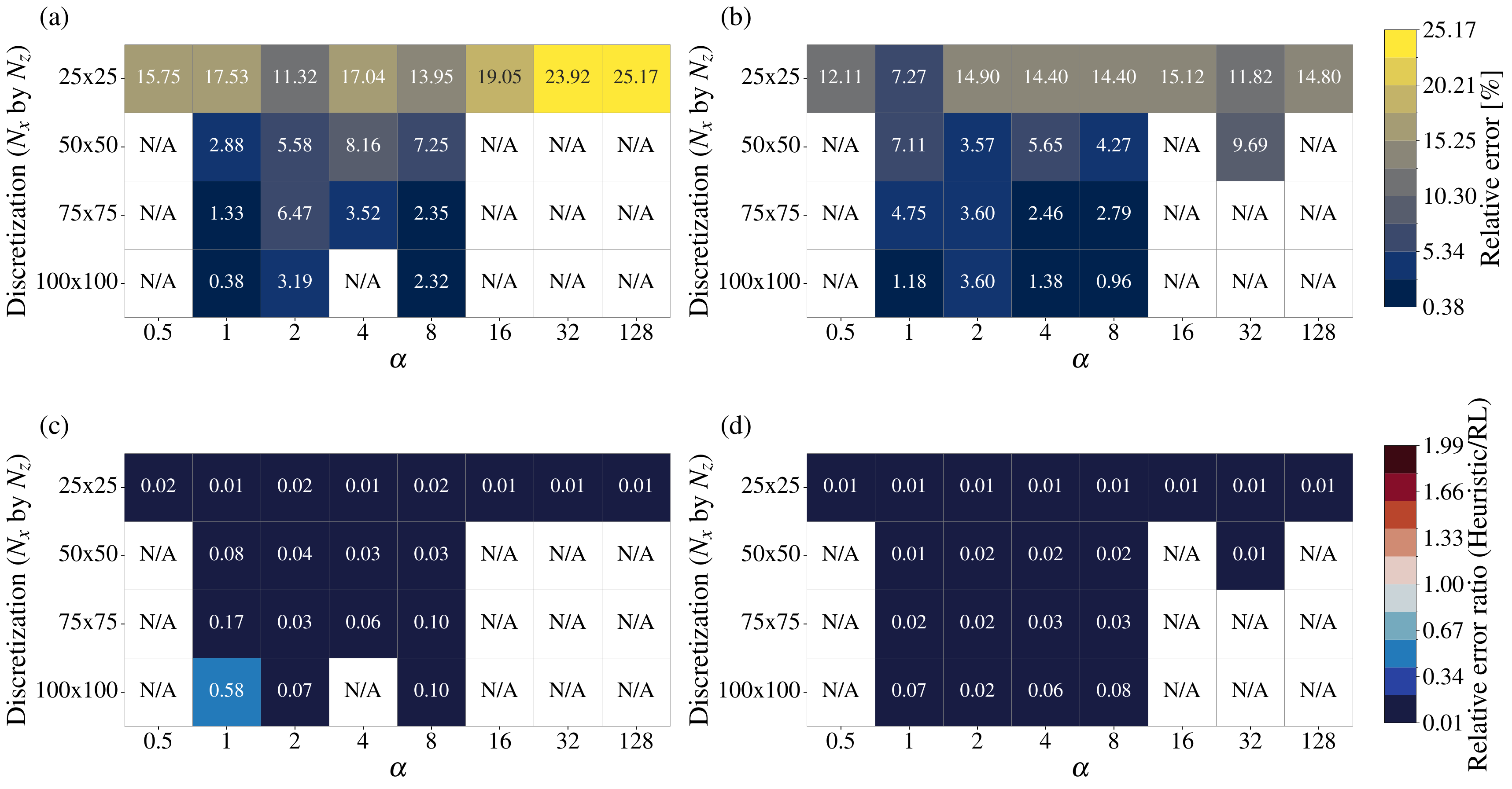}
\caption{Relative error of the spatial average of the time-integrated slip rate of the RL-based method (top row) for (a) $L_x  = L_z = 3$ [km] fault and (b) $L_x  = L_z = 5$ [km] fault and ratio of relative error of the heuristic and the RL-based method for (c) $L_x  = L_z = 3$ [km] fault and (d) $L_x  = L_z = 5$ [km]. The x-axis depicts the speed-accuracy trade-off parameter $\alpha$ (see~\cref{eq: reward}), and the y-axis depicts the fault discretization ($N_x$ by $N_z$). Note that the white boxes with the ``N/A'' text indicate that scenarios in the RL-based integrator did not converge.}\label{fig: accuracy_comp}
\end{figure*}

For a more extensive evaluation, the accuracy is evaluated for the same scenarios displayed in~\cref{sec: accuracy}.~\cref{fig: accuracy_comp} displays the error for the cumulative slip for the RL-based method relative to a constant time stepper in the top row via the color bar, and in the bottom row, its performance relative to the heuristic-based method is shown by the color bar. Note that we take the cumulative slip for the error measurement as it presents the upper bound for errors, given that it is an integral quantity, thus presenting the worst case.~\cref {fig: accuracy_comp}~(a) shows that the RL-based method maintains a good accuracy as the average relative error is approximately 9.85 \%. The highest relative error in all convergent scenarios is 25.17 \%, which corresponds to a deviation of roughly 50 [mm]. Furthermore,~\cref{fig: accuracy_comp} (b) demonstrates that the RL-based method can actually improve accuracy even for a fault configuration not included in the training set, with an average relative error of 7.42 \%. The above error might appear high, but it is not unexpected, given that the gain in calculation time with the RL time stepper is achieved by selecting larger time steps. Indeed, the relative error tends to decrease as the mesh is refined, likely because the timestep is reduced, thereby reducing error accumulation. 

~\cref{fig: accuracy_comp} (c) and (d) show that the relative performance of accuracy varies very little across scenarios, with essentially a value of on the order of one-tenth. On average, for scenarios in~\cref{fig: accuracy_comp}~(c), the RL-based method is 0.074 times as accurate. For a fault configuration of 5 [km] by 5 [km], the RL-based approach is, on average, 0.023 times as accurate as the heuristic-based approach. These results demonstrate that the RL-based method offers a speed advantage at the expense of accuracy. Even in scenarios where the RL-based method is less accurate than the heuristic method. This is not surprising given the nature of the underlying numerical integration scheme, which naturally accumulates more error for larger time step sizes.

\subsubsection{Proficiency}
\label{sec: proficiency}
The preceding analyses reveal that the RL-based adaptive time-stepping method achieves its primary objective of accelerating earthquake fault simulations, consistently outperforming both heuristic and PI adaptive steppers by factors ranging from 4$\times$ to 70$\times$. This remarkable computational efficiency emerges from the method's learned ability to exploit the underlying physics of stick-slip dynamics, maintaining nearly constant time steps during slip events while suddenly increasing step sizes during mechanically stable sticking phases. Such state-dependent behavior contrasts sharply with the heuristic approach's greedy maximization strategy, which repeatedly takes overly optimistic steps that fail to converge, and with the PI stepper's assumption of error regularity, which proves incompatible with the nonsmooth dynamics inherent to frictional systems.

The computational advantages, however, come at the cost of reduced accuracy in certain integral metrics. The RL method exhibits average relative errors of 9.85\% and 7.42\% in cumulative slip for the 3 km and 5 km fault configurations, respectively, relative to the reference constant-time-step solution. Yet this accuracy metric may overstate the practical impact on solution quality. Cumulative slip, as an integral quantity, represents an upper bound on error accumulation—each time step's local error compounds over the entire simulation duration. In many earthquake-dynamics applications, the instantaneous slip-velocity field carries greater physical significance, as it directly governs seismic wave generation, dynamic stress transfer, and the activation of weakening mechanisms that control rupture propagation.

Indeed, visual inspection of the velocity profiles demonstrates that the RL method captures the essential dynamics with high fidelity despite differences in accumulated slip. The maximum velocity error of approximately 0.03 m/s suggests that the method preserves the critical features of the rupture process while accepting controlled degradation in integral quantities. This selective accuracy reflects the optimization inherent in the reward function design—the RL agent has learned to identify which aspects of the solution most strongly influence convergence and to focus its computational effort accordingly.

The observed trade-off between speed and accuracy appears particularly favorable for typical earthquake simulation use cases. Large-scale parametric studies that explore variations in initial stress conditions, friction parameters, or fault geometry often require hundreds or thousands of individual simulations. In such contexts, a 10$\times$ speedup that enables exploration of a much broader parameter space may provide greater scientific value than maintaining strict accuracy in cumulative metrics. Similarly, for real-time or near-real-time applications such as earthquake early warning systems or dynamic rupture forecasting, the ability to complete simulations within stringent time constraints outweighs the need for precise long-term slip accumulation. Moreover, the method's successful generalization to the 5 km$\times$5 km fault configuration, which was not included in the training set, demonstrates robustness and suggests that the learned policy captures fundamental aspects of the adaptive time-stepping problem rather than overfitting to specific training scenarios.

Nevertheless, the convergence difficulties observed for extreme values of the speed-accuracy parameter $\alpha$ reveal important limitations. For $\alpha\leq0.5$, the policy's aggressive pursuit of computational speed leads to repeated selection of non-convergent time steps, while for $\alpha\geq16$, an overly conservative approach paradoxically results in convergence failures—though the precise mechanism for this latter behavior warrants further investigation. These boundaries suggest that the learned policy operates within a specific regime of the speed-accuracy trade-off space, beyond which the assumptions encoded during training no longer hold. Future work might address these limitations by modifying reward functions to incorporate additional penalties for non-convergence or exploring different reward function forms.

\subsubsection{Transfer learning}
\label{sec: transfer_learning}
\begin{figure}[!t]
\centering
\includegraphics[width=0.95\textwidth,height=0.75\textheight,keepaspectratio]{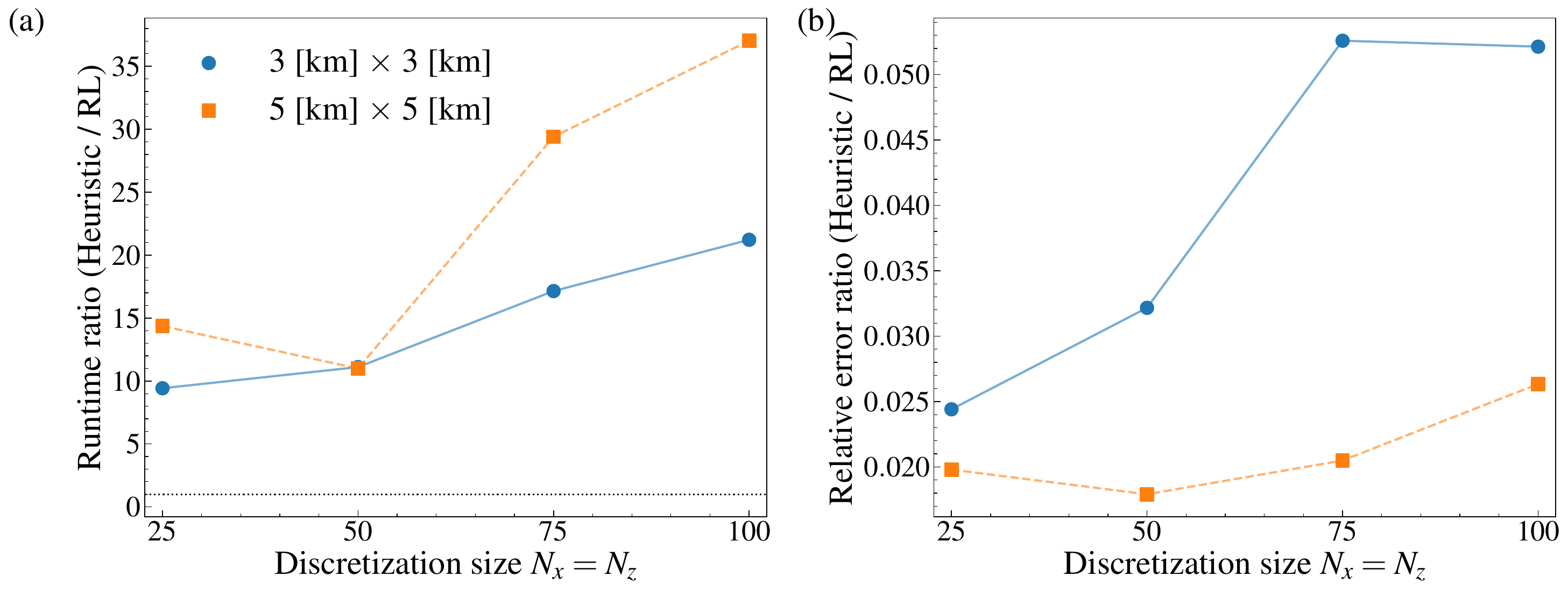}
\caption{Transfer learning results for the speed-accuracy trade-off parameter $\alpha=8$ (see~\cref{eq: reward}) for the runtime ratio (a) and relative error ratio (b) against the fault size discretization for both the two fault sizes.}\label{fig: transfer_learning}
\end{figure}

While the previous sections demonstrated that the proposed RL-based adaptive time stepper offers faster speed than the heuristic-based and PI methods, it also showed relatively low accuracy. To address this issue further, we seek to train the TQC network on a finer discretization (transfer learning) to improve computational speed and accuracy. The trained TQC network for $\alpha=8$ is further trained on a new scenario (transfer learning). The new scenario is for a fault with $L_x=L_z=3$ [km] configuration and a 30-by-30 element discretization.

\cref{fig: transfer_learning} displays the runtime ratio (heuristic/RL) (a) and the relative error ratio (b) against the mesh discretization for both the 3 [km] by 3 [km] and 5 [km] by 5 [km] scenarios. Notably, as expected from transfer learning, the runtime ratio increased across all scenarios. For example, at a discretization of 100 for the 5 [km] by 5 [km] scenario, the runtime ratio following transfer learning is roughly 35, whereas prior to transfer learning it was 23 (see~\cref{fig: speed_comp}). This represents a 1.5x improvement. Notably,~\cref{fig: transfer_learning} (b) shows that while at the coarser discretizations the ratio has improved compared to the originally trained network, it now does not have as large of relative error ratio at the finer discretizations (see~\cref{fig: accuracy_comp} (c) and (d)). This likely is indicative that presumed reward structure~\cref{eq: reward} biases towards speed and that the errors of 2 \% could be too demanding.

\section{Conclusions}
\label{Conclusions}
Our study demonstrates that Reinforcement Learning (RL), specifically using the Truncated Quantile Critics (TQC) algorithm with a continuous action space, can effectively learn adaptive time-stepping policies for nonsmooth dynamical systems, including but not limited to sliding mode controllers, electrical circuits, and frictional interfaces. More specifically, the RL-based adaptive integration exhibited a pronounced improvement in computational speed for the sliding mode controller and the electrical circuit. Furthermore, in an isolated seismic fault system with set-valued Coulomb friction, the method generalizes well to different mesh discretizations and fault configurations, even when trained on a single element for a single slip scenario. This results in an efficient training procedure for rapid deployment in various nonsmooth dynamical systems.

The RL-based approach outperforms traditional heuristic and PI adaptive time-stepping methods in terms of computational speed but maintains sufficient accuracy. This is particularly interesting for complex spatiotemporal scenarios, such as the seismic fault model presented here, where the RL policy adapts to rapid changes in system dynamics more effectively than conventional methods. In particular, our approach achieved a notable speed-up compared to the more conventional heuristic integration scheme. However, the RL-based method failed to converge when certain fault discretizations were used. Finally, it was found that transfer learning does improve computational runtimes.

It is worth noting that the neural network environment requires only information on internal energy and the potential maximum of the state variables, suggesting it should generalize well to many other systems. Future work could explore a broader range of frictional systems, contact mechanics, and elastoplasticity. Furthermore, in the context of constitutive modeling of elastoplasticity, the formulations could be expressed as variational inequalities~\cite{krabbenhoft2007formulation,krabbenhoft2007interior} or, equivalently, a normal inclusion~\cite{guillet2024semi}, allowing for the application of RL-based adaptive time integration. Beyond purely mechanical systems, the RL-based adaptive time integration has interesting applications in the nonsmooth implementation of more complicated sliding mode controllers, such as twisting and super-twisting controllers~\cite{b:SMC_Fridman}, as well as in nonsmooth electrical current modeling. Finally, investigating the impact of different reward function designs, observational space, and hyperparameter choices on the learning process could further enhance the performance of RL-based adaptive time-stepping.

Another promising direction is to let RL tune the internal parameters of classical PI step-size controllers, rather than only selecting the timestep. These controllers expose stability and damping gains that are usually hand-tuned for each problem. For example,~\cite{gutierrez2025ai} used RL to adapt controller gains and improved performance. The PI controller used here~\cite{arevalo2020software} has similar free hyperparameters that could be updated by an RL policy. This would shift adaptive integration from “learning the next step size” to “learning the behaviour of the integrator itself,” potentially yielding problem-aware methods that automatically specialize to particular classes of nonsmooth dynamics without additional manual tuning.

\backmatter

\bmhead{Acknowledgements}

DMR, DGO, and IS want to acknowledge the European Research Council’s
(ERC) support under the European Union’s Horizon 2020 research and
innovation programme (Grant agreement no. 101087771 INJECT). AS, DGO, and IS want to acknowledge the
Region Pays de la Loire and Nantes Métropole under the Connect Talent
programme (CEEV: Controlling Extreme EVents—Blast: Blas LoAds on
STructures). Additionally, the authors would like to thank the reviewers for their time and insightful comments. Finally, IS would like to acknowledge the Hi! Paris chain on AI IA-SUR.

\section*{Declarations}
The authors declare that they have no known competing financial interests or personal relationships that could have influenced the work reported in this paper.
\section*{Conflict of interest/Competing interests}
The authors declare they have neither conflict of interest nor competing interests
\section*{Data availability}
No datasets were generated or analyzed during the
current study.
\section*{Code availability}
The code accompanying this paper is available at
\href{https://github.com/ERC-INJECT/solve_nivp}{https://github.com/ERC-INJECT/solve\_nivp}.

\section*{Author contribution}
\textbf{David Riley}: Formal analysis, Investigation, Methodology, Software, Validation, Visualization, Writing – original draft. \textbf{Alexandros Stathas}: Methodology, Validation, Supervision, Writing – review \& editing. \textbf{Diego Guti\'errez-Oribio}: Methodology, Validation, Supervision, Writing – review \& editing. \textbf{Ioannis Stefanou:} Conceptualization, Methodology, Validation, Supervision, Project administration, Funding acquisition, Writing – review \& editing.

\newcounter{eqsave}\setcounter{eqsave}{\value{equation}}
\newcounter{figsave}\setcounter{figsave}{\value{figure}}
\newcounter{tabsave}\setcounter{tabsave}{\value{table}}

\begin{appendices}

\section{Proof of finite time convergence for FOSM}
\label{app0}

\counterwithout{equation}{section}
\counterwithout{figure}{section}
\counterwithout{table}{section}
\renewcommand\theequation{\arabic{equation}}
\renewcommand\thefigure{\arabic{figure}}
\renewcommand\thetable{\arabic{table}}

\setcounter{equation}{\value{eqsave}}
\setcounter{figure}{\value{figsave}}
\setcounter{table}{\value{tabsave}}

Following the procedure outlined in~\cite{b:SMC_Fridman} and considering the sliding mode system given in~\cref{eq: sliding mode}, we define a candidate Lyapunov function as 

\begin{equation} \label{eq: lyp_candidate} 
V(w) = U = \frac{1}{2}w^2. 
\end{equation} 

Taking the time derivative of $V(w)$ along the trajectories of the system yields 

\begin{equation} 
\dot{V}(w) = w\dot{w} = w(\phi(t) - k\text{sign}(w)). 
\end{equation} 
Since $ w \text{sign}(w)=|w|$, this expression can be rewritten as 
\begin{equation}
\dot{V}(w) = w\phi(t) - k|w|. 
\end{equation}

As the disturbance $\phi(t)$ is bounded by a constant $D$ (i.e., $\phi(t) = D sin(\omega t)\leq D$), it follows that

\begin{equation} \label{eq: lyp_bound} 
\dot{V}(w) \leq -(k-D)|w|. 
\end{equation} 
Thus, if $k>D$,  $\dot{V}<0~\forall w \neq 0$, and therefore, the system is asymptotically stable.

By substituting $|w|=\sqrt{2V(w)} $ into~\cref{eq: lyp_bound}, we obtain
\begin{equation} 
\dot{V}(w) \leq -(k-D)\sqrt{2V(w)}.
\end{equation} 
Applying separation of variables and integrating yields \begin{equation} \label{eq: int_result} 
\sqrt{V(t)} \leq \sqrt{V(0)} - \frac{(k-D)}{\sqrt{2}}t. \end{equation} 
Using $\sqrt{V(t)}=\frac{1}{\sqrt{2}}|w(t)|$ from the Lyapunov function, it follows that 
\begin{equation} \label{eq: bound}
|w(t)| \leq |w(0)| - (k-D)t,
\end{equation}
which is useful for determining an upper bound for $w(t)$ and identifying the normalization constant of the observation space, $U_c$.

Finally, the settling time $T$ can be contained from \eqref{eq: bound} as 
\begin{equation} 
T \leq \frac{|w(0)|}{k-D},
\end{equation} 
which shows that the state $w(t)$ reaches zero in finite time.

\section{Boundedness of electrical circuit}
\label{app: ec}
We demonstrate boundedness by establishing global asymptotic convergence to the equilibrium. That ensures that the internal energy is always decreasing. Again, we first define the internal energy and candidate Lyapunov function as
\begin{equation} \label{eq: ec_ie}
    V(w) =U = \frac{C_1}{2}V_1^2+\frac{C_2}{2}V_2^2+\frac{L}{2}i^2.
\end{equation}
Taking the time derivative and simplifying, we arrive at
\begin{equation}
    \dot{V}= - \frac{(V_1-V_2)^2}{R}-m|V_1| 
\end{equation}
Since $R > 0$ and $m > 0$ for physical components, $\dot{V} \leq 0\quad\forall~V_1,~V_2,~i \in \mathbb{R}$. This establishes that the Lyapunov function is non-increasing. However, the derivative is only negative semi-definite, as $\dot{V} = 0$ when $V_1 = V_2 = 0$ (regardless of the value of $i$). Therefore, we cannot immediately conclude asymptotic stability using standard Lyapunov theory.

To address this, we invoke LaSalle's invariance principle and follow the procedure from~\cite{Khalil2002}. We need to identify the largest invariant set contained in 
$E = \{(V_1, V_2, i) \in \mathbb{R}^3 : \dot{V} = 0\}$. From our expression for $\dot{V}$, we observe that $\dot{V} = 0$ if and only if both $(V_1 - V_2)^2 = 0$ and $m|V_1| = 0$ simultaneously. This occurs when $V_1 = V_2 = 0$, and since $\dot{V}_2 = 0$, yielding $i=0$ for the set to be invariant. At the point $(V_1, V_2, i) = (0, 0, 0)$, all derivatives vanish and, consequently, the largest invariant set contained in $E$ is the origin $\{(0, 0, 0)\}$.

By LaSalle's invariance principle, all system trajectories converge to this invariant set as $t \rightarrow \infty$. Therefore, the electrical circuit exhibits global asymptotic stability at the origin, which ensures that the system's energy remains bounded for all time.



\end{appendices}


\end{document}